\documentclass{aa}
\usepackage{txfonts}
\usepackage{graphicx}
\usepackage[usenames,dvipsnames]{color}
\usepackage{hyperref}
\usepackage{natbib}
\begin{document}

\title{Hot horizontal branch stars in NGC\,288 -- effects of
  diffusion and stratification on their atmospheric parameters\thanks{Based on
    observations with the ESO Very Large Telescope at Paranal
    Observatory, Chile (proposal ID 71.D-0131)},\thanks{Tables 1 and 2
    are available in electronic form at \url{http://www.aanda.org}},\thanks{The observed abundances
    plotted in Fig.\,\ref{Fig:abu} are only available in electronic form at the
    CDS via anonymous ftp to cdsarc.u-strasbg.fr (130.79.128.5) or via
    \url{http://cdsweb.u-strasbg.fr/cgi-bin/qcat?J/A+A/565/A100}}}
\titlerunning{Hot horizontal-branch stars in NGC\,288 -- diffusion
and stratification effects}
\author{S. Moehler\inst{1,2}
\and S. Dreizler\inst{3}
\and F. LeBlanc \inst{4}
\and V. Khalack \inst{4}
\and G. Michaud\inst{5}
\and J. Richer\inst{5}
\and A. V. Sweigart\inst{6}
\and F. Grundahl\inst{7}
} 
\institute{ European Southern Observatory,
  Karl-Schwarzschild-Str. 2, D 85748 Garching, Germany \email{smoehler@eso.org} 
 \and Institut f\"{u}r Theoretische Physik und
  Astrophysik, Olshausenstra\ss e 40, 24118 Kiel, Germany 
 \and Georg-August-Universit\"at, Institut f\"ur Astrophysik,
  Friedrich-Hund-Platz 1, D 37077 G\"ottingen, Germany \email{dreizler@astro.physik.uni-goettingen.de} 
 \and D\'epartement de Physique et d'Astronomie, Universit\'e de
 Moncton, Moncton, New Brunswick, E1A 3E9, Canada 
   \email{[francis.leblanc, viktor.khalak]@umoncton.ca} 
\and D\'epartement de physique, Universit\'e de Montr\'eal, Montr\'eal,
Qu\'ebec, H3C 3J7, Canada \email{[michaudg, jacques.richer]@umontreal.ca}
 \and NASA Goddard Space Flight Center, Exploration of the Universe
 Division, Code 667, Greenbelt, MD 20771, USA \email{allen.sweigart@gmail.com}  
 \and Stellar Astrophysics Centre,
  Department of Physics \& Astronomy, University of \AA rhus, Ny
  Munkegade 120, 8000 \AA rhus C, Denmark \email{fgj@phys.au.dk}
} 
\date{Received 31 October 2013 / Accepted 11 March 2014}

\abstract
{NGC\,288 is a globular cluster with a well-developed blue horizontal
  branch covering the $u$-jump that indicates the onset of
  diffusion. It is therefore well suited to study the effects of
  diffusion in blue horizontal branch (HB) stars.}
{We compare observed abundances with predictions from stellar evolution
  models calculated with diffusion and from stratified atmospheric
  models. We verify the effect of using stratified model spectra to
  derive atmospheric parameters. In addition, we investigate the nature
  of the overluminous blue HB stars around the $u$-jump. }
{We defined a new photometric index $sz$ from $uvby$ measurements that
  is gravity-sensitive between 8\,000\,K and 12\,000\,K. Using
  medium-resolution spectra and Str\"omgren photometry, we determined
  atmospheric parameters ($T_{\rm eff}$, $\log g$) and abundances for
  the blue HB stars.  We used both homogeneous and stratified model
  spectra for our spectroscopic analyses.}
{The atmospheric parameters and masses of the hot HB stars in NGC\,288
  show a behaviour seen also in other clusters for temperatures
  between 9\,000\,K and 14\,000\,K. Outside this temperature range,
  however, they instead follow the results found for such stars in
  $\omega$\,Cen. The abundances derived from our observations are for
  most elements (except He and P) within the abundance range expected
  from evolutionary models that include the effects of atomic
  diffusion and assume a surface mixed mass of
  10$^{-7}$\,M$_\sun$. The abundances predicted by stratified model
  atmospheres are generally significantly more extreme than observed,
  except for Mg. When effective temperatures, surface gravities, and
  masses are determined with stratified model spectra, the hotter
  stars agree better with canonical evolutionary predictions.  }
{ Our results show definite promise towards solving the long-standing problem of
  surface gravity and mass discrepancies for hot HB stars, but much work is still needed to arrive at a self-consistent solution.}
\maketitle
\color{black}
\section{Introduction}
\label{sec:intro}
Low-mass stars that burn helium in a core of about 0.5\,$M_\sun$ and
hydrogen in a shell populate a roughly horizontal region in the optical
colour-magnitude diagrams of globular clusters, which has earned them
the name horizontal branch ({\bf HB}) stars 
\citep{ten27}. The hot (or blue) HB stars near an effective
  temperature of 11\,500\,K are of special interest because they
  exhibit a number of intriguing phenomena associated with the onset
  of diffusion.

A large photometric survey of many globular clusters by \citet{grca99}
demonstrated that the Str\"omgren $u$-brightness of blue HB stars
suddenly increases near 11\,500\,K.  This $u$-jump is attributed to a
sudden increase in the atmospheric metallicity of the blue HB stars to
super-solar values that is caused by the {\em radiative levitation of
  heavy elements}.  A similar effect can be seen in broad-band $U,
U-V$ photometric data \citep[G1]{fpfd98}.  \citet{beco99,beco00} and
\citet{mosw00} confirmed with direct spectroscopic evidence that the
atmospheric metallicity does indeed increase to solar or super-solar
values for HB stars hotter than the $u$-jump. A list of earlier
observations of this effect can be found in \citet{moeh01}. Later
studies include \citet[M\,3, M\,13, M\,15, M\,68, M\,92, and
  NGC\,288]{behr03}, \citet[NGC\,1904]{fare05}, and
\citet[NGC\,2808]{pare06}. These findings also helped to understand
the cause of the low-gravity problem: \citet{crro88} and
\citet{mohe95,mohe97} found that hot horizontal branch stars -- when
analysed with model spectra of the same metallicity as their parent
globular cluster -- show significantly lower surface gravities than
expected from evolutionary tracks. Analysing them instead with more
appropriate metal-rich model spectra reduces the discrepancies
considerably \citep{mosw00}. The more realistic stratified model
atmospheres of \citet{hulh00} and \citet{lmhh09} reduce the
discrepancies in surface gravity even more (see below for more
details).  Along with the enhancement of heavy metals, a decrease in
the helium abundance by mass $Y$ is observed for stars hotter than
$\approx$11\,500\,K, while cooler stars have normal helium abundances
within the observational errors. The helium abundance for these hotter
stars is typically between 1 and 2 dex smaller than the solar value
(e.g. \citealt{behr03}). A trend of the helium abundance relative to
$T_{\rm eff}$ was discussed by \citet{momo09,movi12}. Finally, blue HB
stars near $\approx$11\,500\,K show a sudden drop in their rotation
rates \citep{perc95,bedj00,behr03}, and in some globular clusters
(e.g., M\,13) a gap in their HB distribution.

The possibility of {\em radiative levitation of heavy elements} and
{\em gravitational settling of helium} in HB stars had been predicted
long ago by \citet{miva83}, but the discovery of its very sudden onset
near 11\,500\,K was a complete surprise. \citet{qcmr09} have shown
that helium settling in HB stars cooler than $\approx$11\,500\,K is
hampered by meridional flow. In stars hotter than this threshold,
helium can settle and the superficial convection zone disappears,
  so that diffusion might occur in superficial regions of these
stars.  Recent evolutionary models of HB stars that include atomic
diffusion calculated by \citet{miri08,miri11}
can reasonably well reproduce the abundance anomalies of several
elements observed in blue HB stars. However, these models do not treat
the atmospheric region in detail.  Instead, they assume an outer
superficial mixed zone of approximately 10$^{-7}$M$_\sun$ below which
separation occurs.

The detection of vertical stratification of some elements, especially
iron, in the atmospheres of blue HB stars lends additional support to
the scenario of atomic diffusion being active in these regions
\citep{klbw07,klbw08,khlb10}, but is at variance with the mixed zone
introduced by \citet{miri08,miri11}.  \citet{hulh00} and
\citet{lmhh09} presented stellar atmosphere models of blue HB stars
that include the effect of vertical abundance stratification on the
atmospheric structure.  These models estimate the vertical
stratification of the elements caused by diffusion by assuming that an
equilibrium solution (i.e. giving a nil diffusion-velocity) for each
species is reached.  Assuming a sudden onset of atomic diffusion near
11\,500\,K, these models predict photometric jumps and gaps
(\citealt{grca99}, G1 in \citealt{fpfd98}) consistent with
observations (see \citealt{lehk10} for more details).  The photometric
changes, with respect to chemically homogeneous atmosphere models,
are due to the modification of the atmospheric structure caused by the
abundance stratification.

\begin{figure*}[ht]
\includegraphics[height=\textwidth,angle=270]{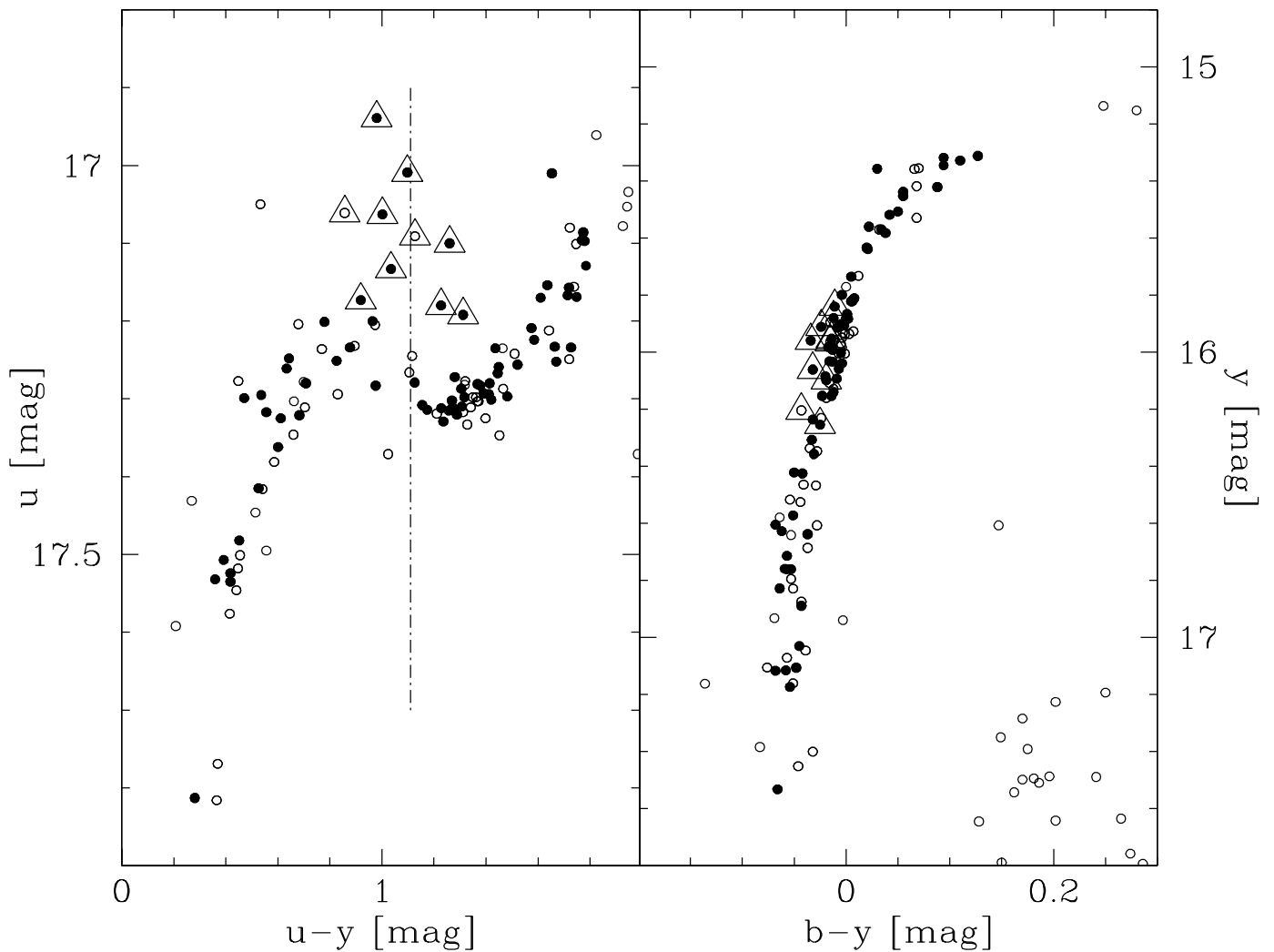}
\caption[]{Blue HB stars in NGC\,288 as observed by
  \citet{grca99}. The subsample, for which GIRAFFE spectra were
  observed, is marked by filled circles. The dot-dashed line in the
  left part marks the $u-y$ colour of the $u$-jump. Stars on the red
  side of the line do not show evidence for radiative levitation,
  while stars on the blue side do. The triangles mark overluminous
  stars around the $u$-jump (from red to blue: 122, 103, 127, 101,
  146, 142, 100, 183).}\label{fig:cmd_hb}
\end{figure*}

The $u$-jump described above can be clearly seen in the
colour-magnitude diagram of NGC\,288 shown in Fig.\,\ref{fig:cmd_hb},
where one also finds a group of stars with large (and unexplained)
scatter in their $u$ magnitudes and ($u-y$) colours around the jump
region (triangles in Fig.\,\ref{fig:cmd_hb}). With maximum errors of
0\fm008 in $u$ and 0\fm003 in $y$ it is unlikely that their positions
are caused by photometric errors. The evolutionary status of these
stars is unclear.  While their bright $u$ magnitudes are suggestive of
radiative levitation, the effect would have to be extreme and some of
them appear to lie on the cool side of the $u$-jump.  Similar groups
of unexpectedly bright stars can be found in other globular clusters
(see M\,2 and M\,92 in \citealt{grca99}), with NGC\,288 presenting the
most pronounced case. We discuss these stars in Sect.\,\ref{sec:lum}
in more detail.

A colour spread along the red giant branch in NGC\,288 first reported
by \citet{yogr08} was identified as a split by \citet{role11}, which
was confirmed by \citet{cabr11} and \citet{momi13}. In their
excellent review on second and third parameters to explain the
horizontal branch morphology, \citet{grca10} estimated that a range in
helium abundance of $\Delta Y$~=~0.012 would explain the temperature
range of the horizontal branch in NGC\,288. This is consistent with
the helium range found more recently from the analysis of
main-sequence photometry by \citet{pimi13}. A variation in helium this
small is unfortunately too small to be detected by our analysis.
\onltab{
\begin{table*}[!h]
\caption[]{Target coordinates and photometric data.\label{tab:targ_phot}}
\begin{tabular}{rlllllll}
\hline
\hline
ID & $\alpha_{2000}$ & $\delta_{2000}$ & $y$ & $b-y$ & c$_1$ & m$_1$\\
\hline
 52 & 00:52:50.77 & $-$26:38:02.9 & 15.357$\pm$0.001 & $+$0.030$\pm$0.003 & $+$1.301$\pm$0.005 & $+$0.131$\pm$0.005\\
 55 & 00:52:47.43 & $-$26:33:14.2 & 15.318$\pm$0.002 & $+$0.094$\pm$0.002 & $+$1.253$\pm$0.004 & $+$0.122$\pm$0.004\\ 
 60 & 00:52:38.59 & $-$26:37:05.1 & 15.328$\pm$0.003 & $+$0.110$\pm$0.004 & $+$1.202$\pm$0.008 & $+$0.118$\pm$0.007\\ 
 61 & 00:52:42.80 & $-$26:36:39.2 & 15.312$\pm$0.001 & $+$0.127$\pm$0.002 & $+$1.151$\pm$0.004 & $+$0.121$\pm$0.003\\ 
 63 & 00:52:47.08 & $-$26:35:25.0 & 15.345$\pm$0.002 & $+$0.094$\pm$0.003 & $+$1.264$\pm$0.005 & $+$0.119$\pm$0.005\\ 
 70 & 00:52:45.40 & $-$26:35:21.5 & 15.438$\pm$0.002 & $+$0.055$\pm$0.003 & $+$1.288$\pm$0.004 & $+$0.133$\pm$0.004\\
 72 & 00:52:53.92 & $-$26:38:45.6 & 15.421$\pm$0.002 & $+$0.088$\pm$0.004 & $+$1.234$\pm$0.008 & $+$0.125$\pm$0.008\\
 74 & 00:52:35.47 & $-$26:34:24.8 & 15.453$\pm$0.001 & $+$0.055$\pm$0.002 & $+$1.283$\pm$0.005 & $+$0.133$\pm$0.003\\ 
 79 & 00:52:42.68 & $-$26:34:50.2 & 15.507$\pm$0.002 & $+$0.050$\pm$0.003 & $+$1.307$\pm$0.004 & $+$0.135$\pm$0.005\\
 81 & 00:52:40.77 & $-$26:33:47.7 & 15.569$\pm$0.001 & $+$0.034$\pm$0.002 & $+$1.288$\pm$0.003 & $+$0.137$\pm$0.003\\
 83 & 00:52:32.07 & $-$26:35:46.6 & 15.560$\pm$0.002 & $+$0.022$\pm$0.002 & $+$1.266$\pm$0.007 & $+$0.139$\pm$0.004\\
 86 & 00:52:52.51 & $-$26:34:29.7 & 15.582$\pm$0.001 & $+$0.038$\pm$0.002 & $+$1.284$\pm$0.006 & $+$0.136$\pm$0.004\\
 88 & 00:52:52.10 & $-$26:34:12.1 & 15.639$\pm$0.002 & $+$0.021$\pm$0.003 & $+$1.260$\pm$0.007 & $+$0.131$\pm$0.005\\
 89 & 00:52:37.88 & $-$26:36:35.4 & 15.518$\pm$0.002 & $+$0.042$\pm$0.002 & $+$1.242$\pm$0.006 & $+$0.134$\pm$0.003\\
 90 & 00:52:46.64 & $-$26:39:03.1 & 15.634$\pm$0.002 & $+$0.020$\pm$0.004 & $+$1.253$\pm$0.011 & $+$0.131$\pm$0.009\\
 96 & 00:52:37.64 & $-$26:31:13.5 & 15.735$\pm$0.004 & $+$0.005$\pm$0.004 & $+$1.238$\pm$0.010 & $+$0.134$\pm$0.007\\
 99 & 00:52:29.92 & $-$26:36:07.4 & 15.799$\pm$0.002 & $-$0.004$\pm$0.003 & $+$1.182$\pm$0.009 & $+$0.133$\pm$0.007\\
100 & 00:52:38.68 & $-$26:35:58.6 & 15.959$\pm$0.002 & $-$0.034$\pm$0.002 & $+$0.858$\pm$0.004 & $+$0.112$\pm$0.004\\
101 & 00:53:04.16 & $-$26:38:29.8 & 15.911$\pm$0.005 & $-$0.024$\pm$0.009 & $+$0.938$\pm$0.011 & $+$0.116$\pm$0.015\\
102 & 00:52:51.47 & $-$26:36:26.0 & 15.815$\pm$0.001 & $+$0.007$\pm$0.002 & $+$1.205$\pm$0.004 & $+$0.128$\pm$0.003\\
103 & 00:52:35.45 & $-$26:39:11.0 & 15.840$\pm$0.003 & $-$0.011$\pm$0.004 & $+$1.063$\pm$0.010 & $+$0.115$\pm$0.007\\
106 & 00:53:02.55 & $-$26:35:32.9 & 15.822$\pm$0.002 & $+$0.005$\pm$0.003 & $+$1.176$\pm$0.008 & $+$0.127$\pm$0.005\\
107 & 00:52:32.39 & $-$26:36:30.2 & 15.810$\pm$0.002 & $+$0.008$\pm$0.002 & $+$1.173$\pm$0.009 & $+$0.126$\pm$0.004\\
111 & 00:53:05.31 & $-$26:32:45.2 & 15.982$\pm$0.003 & $-$0.017$\pm$0.004 & $+$1.084$\pm$0.005 & $+$0.136$\pm$0.005\\
113 & 00:52:39.48 & $-$26:36:45.7 & 15.866$\pm$0.001 & $+$0.001$\pm$0.002 & $+$1.161$\pm$0.006 & $+$0.125$\pm$0.004\\ 
114 & 00:52:38.68 & $-$26:37:49.6 & 15.905$\pm$0.004 & $-$0.002$\pm$0.006 & $+$1.132$\pm$0.015 & $+$0.126$\pm$0.011\\
115 & 00:52:45.24 & $-$26:37:55.1 & 15.881$\pm$0.001 & $+$0.000$\pm$0.002 & $+$1.170$\pm$0.006 & $+$0.125$\pm$0.005\\
118 & 00:52:55.50 & $-$26:35:08.2 & 15.902$\pm$0.001 & $-$0.005$\pm$0.002 & $+$1.158$\pm$0.008 & $+$0.124$\pm$0.003\\
119 & 00:52:56.84 & $-$26:33:44.8 & 15.883$\pm$0.001 & $+$0.002$\pm$0.002 & $+$1.159$\pm$0.007 & $+$0.123$\pm$0.004\\ 
120 & 00:53:01.85 & $-$26:37:53.2 & 15.913$\pm$0.004 & $-$0.008$\pm$0.006 & $+$1.136$\pm$0.007 & $+$0.128$\pm$0.010\\
122 & 00:52:37.78 & $-$26:39:31.6 & 15.880$\pm$0.002 & $-$0.012$\pm$0.004 & $+$1.094$\pm$0.011 & $+$0.127$\pm$0.009\\ 
127 & 00:52:39.32 & $-$26:34:31.6 & 15.953$\pm$0.002 & $-$0.014$\pm$0.002 & $+$1.047$\pm$0.004 & $+$0.111$\pm$0.004\\ 
142 & 00:52:50.55 & $-$26:36:49.8 & 16.061$\pm$0.002 & $-$0.032$\pm$0.002 & $+$0.876$\pm$0.005 & $+$0.111$\pm$0.004\\
143 & 00:52:52.77 & $-$26:34:53.0 & 16.003$\pm$0.002 & $-$0.005$\pm$0.003 & $+$1.084$\pm$0.006 & $+$0.119$\pm$0.006\\
145 & 00:52:40.58 & $-$26:32:48.6 & 15.981$\pm$0.002 & $-$0.013$\pm$0.002 & $+$1.108$\pm$0.004 & $+$0.124$\pm$0.003\\
146 & 00:52:43.35 & $-$26:37:56.4 & 16.098$\pm$0.002 & $-$0.019$\pm$0.004 & $+$0.878$\pm$0.005 & $+$0.107$\pm$0.006\\
147 & 00:52:37.26 & $-$26:36:46.9 & 16.033$\pm$0.002 & $-$0.013$\pm$0.003 & $+$1.054$\pm$0.007 & $+$0.127$\pm$0.004\\
149 & 00:52:33.14 & $-$26:33:44.6 & 16.084$\pm$0.003 & $-$0.020$\pm$0.004 & $+$1.044$\pm$0.009 & $+$0.122$\pm$0.005\\
151 & 00:52:48.32 & $-$26:32:57.6 & 16.032$\pm$0.002 & $-$0.016$\pm$0.003 & $+$1.074$\pm$0.004 & $+$0.131$\pm$0.004\\
154 & 00:52:54.89 & $-$26:37:11.7 & 16.058$\pm$0.001 & $-$0.007$\pm$0.002 & $+$1.054$\pm$0.007 & $+$0.112$\pm$0.004\\
156 & 00:52:50.53 & $-$26:35:12.0 & 16.039$\pm$0.003 & $-$0.004$\pm$0.005 & $+$1.060$\pm$0.007 & $+$0.112$\pm$0.009\\
157 & 00:52:53.76 & $-$26:39:08.7 & 15.992$\pm$0.004 & $-$0.013$\pm$0.007 & $+$1.073$\pm$0.011 & $+$0.123$\pm$0.013\\
167 & 00:52:46.42 & $-$26:34:07.7 & 16.093$\pm$0.002 & $-$0.009$\pm$0.003 & $+$1.039$\pm$0.006 & $+$0.112$\pm$0.006\\
169 & 00:52:46.50 & $-$26:31:30.7 & 16.153$\pm$0.002 & $-$0.023$\pm$0.003 & $+$0.957$\pm$0.006 & $+$0.119$\pm$0.005\\
176 & 00:52:48.70 & $-$26:34:00.7 & 16.140$\pm$0.002 & $-$0.012$\pm$0.003 & $+$0.984$\pm$0.006 & $+$0.113$\pm$0.006\\
179 & 00:52:48.17 & $-$26:35:19.9 & 16.236$\pm$0.003 & $-$0.032$\pm$0.006 & $+$0.844$\pm$0.009 & $+$0.108$\pm$0.011\\
180 & 00:52:50.91 & $-$26:36:09.4 & 16.153$\pm$0.002 & $-$0.014$\pm$0.003 & $+$0.975$\pm$0.004 & $+$0.111$\pm$0.004\\
183 & 00:52:39.91 & $-$26:37:23.8 & 16.254$\pm$0.002 & $-$0.025$\pm$0.004 & $+$0.784$\pm$0.011 & $+$0.105$\pm$0.009\\
187 & 00:52:44.71 & $-$26:35:31.4 & 16.307$\pm$0.003 & $-$0.033$\pm$0.005 & $+$0.843$\pm$0.008 & $+$0.116$\pm$0.009\\
195 & 00:52:27.41 & $-$26:35:58.8 & 16.422$\pm$0.004 & $-$0.050$\pm$0.005 & $+$0.685$\pm$0.008 & $+$0.122$\pm$0.008\\
196 & 00:52:44.29 & $-$26:35:53.2 & 16.357$\pm$0.002 & $-$0.031$\pm$0.002 & $+$0.760$\pm$0.004 & $+$0.105$\pm$0.004\\
199 & 00:52:55.57 & $-$26:32:58.7 & 16.425$\pm$0.002 & $-$0.042$\pm$0.002 & $+$0.726$\pm$0.005 & $+$0.113$\pm$0.004\\
212 & 00:52:59.33 & $-$26:39:00.4 & 16.605$\pm$0.005 & $-$0.068$\pm$0.009 & $+$0.577$\pm$0.013 & $+$0.135$\pm$0.017\\
213 & 00:52:42.83 & $-$26:31:06.8 & 16.627$\pm$0.005 & $-$0.062$\pm$0.005 & $+$0.564$\pm$0.008 & $+$0.128$\pm$0.007\\
\hline
\end{tabular}
\end{table*}

\setcounter{table}{0}
\begin{table*}
\caption[]{Target coordinates and photometric data (continued)}
\begin{tabular}{rllllll}
\hline
\hline
ID & $\alpha_{2000}$ & $\delta_{2000}$ & $y$ & $b-y$ & c$_1$ & m$_1$\\
\hline
216 & 00:52:54.57 & $-$26:33:20.4 & 16.572$\pm$0.002 & $-$0.051$\pm$0.005 & $+$0.627$\pm$0.011 & $+$0.117$\pm$0.010\\
221 & 00:52:52.31 & $-$26:35:13.7 & 16.638$\pm$0.002 & $-$0.037$\pm$0.003 & $+$0.584$\pm$0.007 & $+$0.105$\pm$0.006\\
228 & 00:52:47.36 & $-$26:37:52.6 & 16.761$\pm$0.002 & $-$0.057$\pm$0.004 & $+$0.495$\pm$0.009 & $+$0.116$\pm$0.008\\
230 & 00:52:24.33 & $-$26:35:23.5 & 16.759$\pm$0.009 & $-$0.059$\pm$0.014 & $+$0.485$\pm$0.029 & $+$0.114$\pm$0.025\\
231 & 00:52:47.01 & $-$26:36:23.0 & 16.714$\pm$0.003 & $-$0.057$\pm$0.004 & $+$0.552$\pm$0.008 & $+$0.115$\pm$0.006\\
240 & 00:52:43.69 & $-$26:35:01.7 & 16.761$\pm$0.003 & $-$0.053$\pm$0.005 & $+$0.534$\pm$0.007 & $+$0.113$\pm$0.008\\
242 & 00:52:45.02 & $-$26:37:35.4 & 16.828$\pm$0.002 & $-$0.064$\pm$0.005 & $+$0.421$\pm$0.007 & $+$0.121$\pm$0.009\\
243 & 00:52:44.02 & $-$26:35:42.2 & 16.889$\pm$0.003 & $-$0.043$\pm$0.003 & $+$0.445$\pm$0.006 & $+$0.105$\pm$0.005\\
275 & 00:52:49.39 & $-$26:35:53.7 & 17.030$\pm$0.006 & $-$0.045$\pm$0.009 & $+$0.391$\pm$0.015 & $+$0.098$\pm$0.016\\ 
288 & 00:52:49.30 & $-$26:38:19.1 & 17.115$\pm$0.003 & $-$0.058$\pm$0.004 & $+$0.358$\pm$0.006 & $+$0.104$\pm$0.006\\
292 & 00:53:00.03 & $-$26:36:32.9 & 17.117$\pm$0.003 & $-$0.068$\pm$0.005 & $+$0.406$\pm$0.008 & $+$0.108$\pm$0.008\\
300 & 00:52:49.11 & $-$26:35:35.1 & 17.106$\pm$0.003 & $-$0.048$\pm$0.007 & $+$0.362$\pm$0.009 & $+$0.100$\pm$0.013\\
304 & 00:52:48.60 & $-$26:33:17.1 & 17.173$\pm$0.003 & $-$0.054$\pm$0.005 & $+$0.307$\pm$0.011 & $+$0.107$\pm$0.009\\
347 & 00:52:50.31 & $-$26:38:30.0 & 17.532$\pm$0.003 & $-$0.066$\pm$0.005 & $+$0.275$\pm$0.009 & $+$0.102$\pm$0.009\\
\hline
\end{tabular}
\end{table*}
}
\section{Observations}\label{sec:obs}
The targets were selected from the Str\"omgren photometry of
\citet{grca99}. We selected 71 blue horizontal-branch-star candidates
(see Fig.\,\ref{fig:cmd_hb}) and 17 red giants. Of the blue HB
candidates three were found to be red HB stars and one had extremely
noisy spectra. We here only discuss the observations of the 67
bona-fide hot horizontal branch stars (see Table\,\ref{tab:targ_phot}
for their coordinates and photometric measurements).

The spectroscopic data were obtained between July 3 and 27, 2003 (date
at the beginning of the night, see Table\,\ref{tab:obs} for details)
in Service Mode using the multi-object fibre spectrograph
FLAMES$+$GIRAFFE \citep{pasq00}, which is
mounted at the UT2 Telescope of the VLT. The fibre systems MEDUSA1 and
MEDUSA2 allow one to observe up to 132 objects simultaneously. We
used the low spectroscopic resolution mode with the spectral ranges
3620 -- 4081\,\AA\ (LR1, $\lambda/\Delta\lambda$ = 8000), 3964 --
4567\,\AA\ (LR2, $\lambda/\Delta\lambda$ = 6400), and 4501 --
5078\,\AA\ (LR3, $\lambda/\Delta\lambda$ = 7500).  GIRAFFE had a
2k$\times$4k EEV CCD chip (15$\mu m$ pixel size), with a gain of
0.54\,e$^-$\,ADU$^{-1}$ and a read-out-noise of 3.2\,e$^-$.  Each night
four screen flat-fields, five bias, and one ThAr wavelength
calibration frame were observed. In addition to the daytime ThAr
spectra that covered all fibres, we also observed simultaneous ThAr
spectra during the science observations in five of the fibres.  Dark
exposures were not necessary, because the dark current of
2\,e$^-$\,h$^{-1}$\,pixel$^{-1}$ is negligible.

\onltab{
\begin{table*}
\caption{Observing times, conditions, and setups\label{tab:obs}}
\begin{tabular}{llllrrlr}
\hline
\hline
date & start & seeing & airmass & \multicolumn{2}{c}{Moon} & setup &
exposure\\
     & [UT]  &        &         & illum. & dist. &  & time [s]\\
\hline
2003-07-04 & 08:15:23 & 1\farcs5 & 1.15 & 0.21 & 146\fdg9 & LR1 & 2520 \\
2003-07-04 & 09:06:06 & 1\farcs1 & 1.06 & 0.22 & 147\fdg3 & LR1 & 2520 \\
2003-07-04 & 09:49:14 & 1\farcs1 & 1.02 & 0.22 & 147\fdg7 & LR1 & 2000 \\
2003-07-05$^\ast$ & 07:11:48 & 1\farcs0 & 1.37 & 0.30 & 153\fdg2 & LR1 & 2520 \\
2003-07-05$^\ast$ & 07:56:47 & 1\farcs2 & 1.21 & 0.30 & 153\fdg4 & LR1 & 1432 \\
2003-07-07 & 08:04:26 & 0\farcs9 & 1.15 & 0.52 & 149\fdg7 & LR1 & 2520 \\
2003-07-07 & 08:49:35 & 0\farcs8 & 1.06 & 0.53 & 149\fdg4 & LR1 & 2520 \\
2003-07-07 & 09:33:38 & 0\farcs8 & 1.02 & 0.53 & 149\fdg2 & LR1 & 2520 \\
2003-07-07$^\ast$ & 10:16:53 & 0\farcs8 & 1.00 & 0.53 & 148\fdg9 & LR1 & 832 \\
2003-07-08 & 08:23:19 & 0\farcs6 & 1.10 & 0.63 & 140\fdg2 & LR1 & 2520 \\
2003-07-08 & 09:09:37 & 0\farcs7 & 1.03 & 0.64 & 139\fdg8 & LR1 & 2520 \\
2003-07-08$^\ast$ & 09:53:34 & 0\farcs8 & 1.01 & 0.64 & 139\fdg4 & LR1 & 1417 \\
2003-07-09 & 09:19:31 & 0\farcs7 & 1.02 & 0.75 & 128\fdg2 & LR1 & 2520 \\
2003-07-10 & 09:06:35 & 0\farcs6 & 1.03 & 0.84 & 115\fdg9 & LR3 & 3000 \\
2003-07-10 & 10:01:51 & 1\farcs1 & 1.00 & 0.84 & 115\fdg3 & LR3 & 661 \\
2003-07-20 & 05:36:53 & 0\farcs7 & 1.59 & 0.59 & 30\fdg1 & LR3 & 2520 \\
2003-07-20 & 06:20:22 & 1\farcs0 & 1.40 & 0.59 & 30\fdg3 & LR3 & 623 \\
2003-07-21$^\ast$ & 06:15:02 & 0\farcs6 & 1.34 & 0.50 & 38\fdg0 & LR3 & 2520 \\
2003-07-21 & 06:58:04 & 0\farcs8 & 1.18 & 0.49 & 38\fdg2 & LR3 & 2520 \\
2003-07-21 & 07:47:32 & 1\farcs2 & 1.07 & 0.49 & 38\fdg5 & LR2 & 2520 \\
2003-07-21 & 08:30:33 & 1\farcs5 & 1.02 & 0.49 & 38\fdg7 & LR2 & 2520 \\
2003-07-21 & 10:17:49 & 1\farcs1 & 1.02 & 0.48 & 39\fdg3 & LR2 & 1736 \\
2003-07-22 & 05:03:19 & 0\farcs4 & 1.83 & 0.41 & 46\fdg6 & LR1 & 2520 \\
2003-07-22 & 05:46:25 & 0\farcs4 & 1.47 & 0.40 & 46\fdg9 & LR1 & 2520 \\
2003-07-22 & 06:37:10 & 0\farcs5 & 1.24 & 0.40 & 47\fdg3 & LR2 & 2520 \\
2003-07-22 & 07:22:53 & 0\farcs5 & 1.11 & 0.40 & 47\fdg6 & LR2 & 2520 \\
2003-07-22 & 08:12:00 & 0\farcs5 & 1.04 & 0.40 & 47\fdg9 & LR2 & 2520 \\
2003-07-22$^\ast$ & 10:21:07 & 0\farcs6 & 1.02 & 0.39 & 48\fdg6 & LR2 & 20 \\
2003-07-27 & 07:45:34 & 0\farcs9 & 1.04 & 0.04 & 101\fdg2 & LR1 & 2520 \\
2003-07-28 & 05:59:11 & 0\farcs5 & 1.29 & 0.01 & 111\fdg3 & LR2 & 2520 \\
2003-07-28 & 06:49:13 & 0\farcs5 & 1.13 & 0.01 & 111\fdg9 & LR1 & 2520 \\
\hline
\multicolumn{6}{l}{$^\ast$: not used in final analysis, usually due to
S/N}\\ 
\end{tabular}
\end{table*}
}

\section{Data reduction}\label{sec:reduction}
The spectroscopic data were reduced using the girBLDRS software
(\url{http://girbldrs.sourceforge.net/}, version 1.10) and ESO MIDAS (see
\citealt{drews2005} for details). The two-dimensional bias and
flat-field frames were averaged for each night. The averaged
flat-fields were used to determine the positions and widths of the
fibre spectra. Portions of 64-pixels of each fibre spectrum were
averaged and fitted with a point spread function ({\bf PSF}). A
polynomial fit to the PSF parameters then provides the PSF for the
whole frame, which is used for all extractions later on. 
  One-dimensional flat-field spectra extracted using Horne's method
\citep{horn86} are used to correct the extracted science
  spectra for spectrograph and detector signature. Because the
    flat-field spectra are not normalised, they still show the
  spectral signature of the flat-field lamp, which varies only slowly
  with wavelength, however.

The daytime Th-Ar wavelength calibration spectra were extracted for each fibre
via a simple sum, the spectral lines were localised and fitted by an
analytical model. The presence of simultaneous ThAr spectra in the
science frames allowed us to refine the localisation and wavelength
calibration of the science data by correcting for residual wavelength
shifts between daytime and nighttime observations. The refinement was
repeated until the difference between the observed ThAr positions and
the laboratory ones was below 0.001\,km\,s$^{-1}$.
The wavelength offsets obtained from the
cross-correlation were then linearly interpolated across the CCD
(spatial direction), and the interpolated offsets were applied to each
fibre. 

The raw science spectra were bias and flat-field corrected and finally
rebinned to constant wavelength steps. The median of the sky signal
obtained in 18 dedicated fibres was subtracted from the science data.

The spectra for a given star and setup were then normalised as
described in \citet{modr11}, taking care to use only
regions free from strong lines for the continuum definition. This
  normalisation allows us to exclude outlying pixels during the
averaging of the spectra.

\section{Radial velocities}\label{sec:RV}
After the barycentric correction, the observed spectra were first
co-added (without further radial velocity correction) and fitted with
stellar model spectra to obtain a first estimate of their effective
temperatures, surface gravities, and helium abundances (see
Sect.\,\ref{sec:lineprof} for details). The individual spectra of each
star were then cross-correlated (see \citealt{toda79}
  for more details) with the best-fitting synthetic spectrum derived
this way. Only regions of hydrogen or helium lines were selected prior
to the cross-correlation. The peak of the cross-correlation function
was fitted with a Gaussian function to determine the radial velocity
to sub-resolution accuracy. The velocity-corrected spectra were
co-added and fitted with synthetic model atmospheres (see
Sect.\,\ref{sec:param}). In a second step, the best-fit synthetic
spectra were then used to repeat the cross-correlation.  The 1$\sigma$
errors in radial velocity for individual spectra range from
1.8\,km\,s$^{-1}$ to up to 10\,km\,s$^{-1}$ depending on the
quality. The median error is 2.7\,km\,s$^{-1}$.

For each star, we calculated the standard deviation of the radial
velocity measurements of the individual spectra. We then determined
how many of the individual measurements deviate more than 1$\sigma$
from the mean. In none of the stars was this more than 30\%, which
means that the scatter is consistent with a random distribution within
the measurement error. We thus find no evidence for radial velocity
variations that might indicate the presence of close binaries in our
sample. We note that this absence of evidence should not be confused
with an evidence of absence, that is, we cannot say anything about the
frequency of binaries among our target stars, because neither our
method nor our observations were tailored towards the detection of
binaries.  There is no evidence for field contamination in our sample
either.

The radial velocities of the 67 stars range from $-$48.6\,km\,s$^{-1}$
to $-$36.9\,km\,s$^{-1}$ with a median value of
$-$42.6$\pm$2.7\,km\,s$^{-1}$, which is close to the average value of
$-$45.4\,km\,s$^{-1}$ given by \citet{harr96} and to the recently
published values of $-$43.5\,km\,s$^{-1}$ \citep[uncertainty
  1--2\,km\,s$^{-1}$]{szki07}, $-$46.15$\pm$2.55\,km\,s$^{-1}$
\citep{cabr09a}, and $-$45.1$\pm$0.2\,km\,s$^{-1}$ \citep{laki10}.

Each spectrum was corrected to laboratory wavelength.  As an
additional safeguard against outliers, only a subset of pixel values
around the median (11 out of 13 for LR1, 5 out of 7 for LR2, 3 out of
5 for LR3) was averaged. To allow a combination of the three
wavelength ranges into one spectrum for fitting, the LR1 data were
convolved with a Gaussian with a FWHM of 0.48\,\AA\ so that we had the
same resolution for all three regions.

\section{Atmospheric parameters}\label{sec:param}
\subsection{Determination from photometric data}\label{sec:phot_par}
We used the Str\"omgren photometry of \citet{grca99} to derive a
first estimate of the effective temperatures and surface gravities of
stars redwards of the $u$-jump, similar to our work in
\citet{mola03}. As reference we used theoretical colours from Kurucz (ATLAS9,
\citeyear{kuru93}) for metallicities [M/H] = $-$1.0 and $-$1.5. The
metallicity [Fe/H] of NGC\,288 given by \citet{cabr09b} is
$-$1.32. Unfortunately, we have no H$\beta$ photometry for the
stars. Therefore, guided by the equations of \citet{modw85}, we
searched for a combination of colours that provides a rectangular grid
in the $T_{\rm eff}$, $\log g$ plane for the range 8\,000\,K $\le
T_{\rm eff} \le$ 12\,000\,K. We used the definition for $a_0$ from
Moon \& Dworetsky, but in addition, we defined an index $sz$
\begin{equation}
 a_{0} = 1.36 \cdot (b-y) + 0.36 \cdot m_1 + 0.18 \cdot c_1 - 0.2448 
\end{equation}
\begin{equation}
 sz = -0.07 \cdot (b-y) - m_1 + 0.1 \cdot c_1 + 0.1. 
\end{equation}

\begin{figure}[h!]
\includegraphics[height=\columnwidth,angle=270]{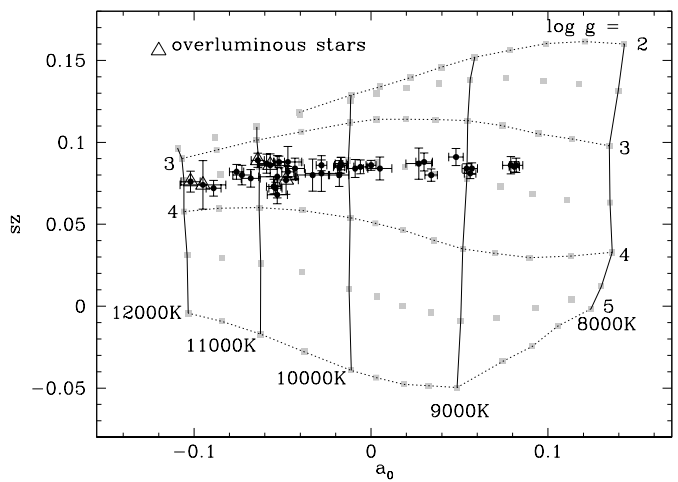}
\caption[]{Theoretical $a_0, sz$ grid as derived from
  \citet{kuru93} colours for [Fe/H] = $-$1.5.  In addition, we
  show the positions of the cool blue HB stars (i.e. redward of the
  vertical line in Fig.\,\ref{fig:cmd_hb}) in this diagram as observed by
  \citet{grca99}.
\label{fig:phot_grid} }
\end{figure}

As one can see from Fig.\,\ref{fig:phot_grid} , $a_{0}$ correlates with
$T_{\rm eff}$ and $sz$ correlates with $\log g$ and $T_{\rm eff}$. The photometric
  data were dereddened with $E_{\rm {B-V}}$ = 0\fm03
\citep{cagr00}, using $E_{\rm b-y}$ = 0.75$\cdot E_{\rm {B-V}}$.

\begin{table*}[!ht]
\caption[]{Temperatures and surface gravities for stars redder than
  the $u$-jump that do not show evidence of radiative levitation as
  derived from the $a_0$ and $sz$ parameters and from line profile
  fits for an assumed metallicity of [M/H] = $-$1.5.  The errors for
  the photometric results are derived
  from the errors of $a_0$ and $sz$ in combination with the errors of
  the fits and an uncertainty of 0\fm02
for the reddening (see Sect.\,\ref{sec:phot_par} for details). The
errors for the spectroscopic results are derived from the $\chi^2$ fit
(see Sect.\,\ref{sec:lineprof} for details) and are only statistical
errors. In addition, we list the helium abundance as determined with
LINFOR (see Sect.\,\ref{sec:abu} for details).}\label{tab:par_phot}
\begin{tabular}{rrrrrrrr}
\hline
\hline
ID & \multicolumn{4}{c}{photometric parameters} & \multicolumn{3}{c}{spectroscopic parameters}\\
 & $a_0$ & $sz$ & $T_{\rm eff}$ & $\log g$ & $T_{\rm eff}$ & $\log g$
& ${\rm \log{(\frac{n_{He}}{n_{H}}}})_{\rm LINFOR}$\\
 & [mmag] & [mmag] & [K] &  & [K] & \\
\hline
 52 &  $+$48$\pm$\ \ 4 &  91$\pm$\ \ 5 &  9100$\pm$380 & 3.39$\pm$0.11 & 8240$\pm$\ \ 20 &  2.87$\pm$0.02 & \\ 
 55 & $+$124$\pm$\ \ 4 &  91$\pm$\ \ 4 &  8100$\pm$260 & 3.08$\pm$0.15 & \multicolumn{2}{c}{ } & \\ 
 60 & $+$134$\pm$\ \ 6 &  89$\pm$\ \ 7 &  8000$\pm$260 & 3.05$\pm$0.19 & \multicolumn{2}{c}{ } & \\ 
 61 & $+$151$\pm$\ \ 3 &  79$\pm$\ \ 3 &  7900$\pm$210 & 3.09$\pm$0.16 & \multicolumn{2}{c}{ } & \\ 
 63 & $+$125$\pm$\ \ 5 &  95$\pm$\ \ 5 &  8100$\pm$270 & 3.02$\pm$0.17 & \multicolumn{2}{c}{ } & \\ 
 70 &  $+$82$\pm$\ \ 4 &  86$\pm$\ \ 4 &  8600$\pm$330 & 3.36$\pm$0.12 & 7990$\pm$\ \ 20 &  2.94$\pm$0.03 & \\ 
 72 & $+$114$\pm$\ \ 7 &  86$\pm$\ \ 8 &  8200$\pm$300 & 3.21$\pm$0.18 & \multicolumn{2}{c}{ } & \\ 
 74 &  $+$81$\pm$\ \ 3 &  85$\pm$\ \ 3 &  8600$\pm$320 & 3.37$\pm$0.11 & 7910$\pm$\ \ 10 &  2.94$\pm$0.04 & \\ 
 79 &  $+$79$\pm$\ \ 5 &  86$\pm$\ \ 5 &  8600$\pm$340 & 3.37$\pm$0.13 & 7990$\pm$\ \ 10 &  2.97$\pm$0.03 & \\ 
 81 &  $+$54$\pm$\ \ 3 &  84$\pm$\ \ 3 &  9000$\pm$350 & 3.47$\pm$0.10 & 8210$\pm$\ \ 20 &  2.98$\pm$0.02 & \\ 
 83 &  $+$34$\pm$\ \ 4 &  80$\pm$\ \ 4 &  9300$\pm$400 & 3.57$\pm$0.10 & 8360$\pm$\ \ 30 &  3.02$\pm$0.02 & \\ 
 86 &  $+$57$\pm$\ \ 3 &  84$\pm$\ \ 4 &  8900$\pm$360 & 3.46$\pm$0.10 & 8160$\pm$\ \ 20 &  3.00$\pm$0.02 & \\ 
 88 &  $+$30$\pm$\ \ 5 &  88$\pm$\ \ 6 &  9300$\pm$420 & 3.46$\pm$0.11 & 8390$\pm$\ \ 30 &  2.98$\pm$0.02 & \\ 
 89 &  $+$56$\pm$\ \ 3 &  81$\pm$\ \ 3 &  8900$\pm$360 & 3.51$\pm$0.10 & 8070$\pm$\ \ 20 &  2.93$\pm$0.03 & \\ 
 90 &  $+$27$\pm$\ \ 7 &  87$\pm$10    &  9400$\pm$470 & 3.48$\pm$0.14 & 8450$\pm$\ \ 30 &  3.04$\pm$0.02 & \\ 
 96 &   $+$5$\pm$\ \ 7 &  84$\pm$\ \ 7 &  9700$\pm$500 & 3.54$\pm$0.12 & 8560$\pm$\ \ 30 &  3.07$\pm$0.02 & \\ 
 99 & $-$17$\pm$\ \ 6 &  80$\pm$\ \ 7  & 10100$\pm$510 & 3.61$\pm$0.12 & 9130$\pm$260 &  3.25$\pm$0.15 & $-$0.6 \\ 
102 &   0$\pm$\ \ 3 &  86$\pm$\ \ 3 &  9800$\pm$440 &  3.51$\pm$0.09 & 9290$\pm$320 &  3.38$\pm$0.18 & $-$1.4 \\ 
103 & $-$56$\pm$\ \ 6 &  86$\pm$\ \ 7 & 10900$\pm$590 &  3.42$\pm$0.19 & 9850$\pm$\ \ 70 &  3.34$\pm$0.04 & $-$1.0 \\ 
106 &  $-$8$\pm$\ \ 5 &  84$\pm$\ \ 6 & 10000$\pm$490 &  3.54$\pm$0.11 & 9230$\pm$290 &  3.33$\pm$0.17 & $-$1.3 \\ 
107 &  $-$5$\pm$\ \ 4 &  85$\pm$\ \ 4 &  9900$\pm$460 &  3.52$\pm$0.10 & 9350$\pm$260 &  3.39$\pm$0.15 & $-$1.2 \\ 
111 & $-$52$\pm$\ \ 6 &  68$\pm$\ \ 6 & 10800$\pm$570 &  3.82$\pm$0.12 & 9760$\pm$\ \ 90 &  3.44$\pm$0.05 & $-$1.0 \\ 
113 & $-$17$\pm$\ \ 3 &  85$\pm$\ \ 4 & 10100$\pm$470 &  3.52$\pm$0.11 & 9210$\pm$180 &  3.31$\pm$0.11 & $-$0.8 \\ 
114 & $-$27$\pm$\ \ 9 &  81$\pm$11 & 10300$\pm$600 &  3.58$\pm$0.17 & 9340$\pm$140 &  3.33$\pm$0.08 & $-$1.0 \\ 
115 & $-$16$\pm$\ \ 4 &  86$\pm$\ \ 5 & 10100$\pm$480 &  3.50$\pm$0.12 & 9280$\pm$180 &  3.34$\pm$0.10 & $-$1.1 \\ 
118 & $-$27$\pm$\ \ 3 &  86$\pm$\ \ 3 & 10300$\pm$480 &  3.49$\pm$0.11 & 9440$\pm$\ \ 90 &  3.38$\pm$0.05 & $-$1.1 \\ 
119 & $-$16$\pm$\ \ 3 &  87$\pm$\ \ 4 & 10100$\pm$470 &  3.48$\pm$0.11 & 9430$\pm$170 &  3.39$\pm$0.09 & $-$1.2 \\ 
120 & $-$32$\pm$\ \ 9 &  80$\pm$10 & 10400$\pm$600 &  3.60$\pm$0.17 & 9470$\pm$\ \ 70 &  3.39$\pm$0.04 & $-$1.0 \\ 
122 & $-$47$\pm$\ \ 7 &  77$\pm$\ \ 9 & 10700$\pm$590 &  3.64$\pm$0.17 & 9740$\pm$\ \ 80 &  3.40$\pm$0.04 & $-$1.1 \\ 
127 & $-$63$\pm$\ \ 4 &  89$\pm$\ \ 4 & 11000$\pm$550 &  3.33$\pm$0.18 & 9920$\pm$\ \ 80 &  3.37$\pm$0.04 & $-$0.8 \\ 
143 & $-$42$\pm$\ \ 5 &  84$\pm$\ \ 6 & 10600$\pm$550 &  3.50$\pm$0.16 & 9810$\pm$\ \ 60 &  3.45$\pm$0.03 & $-$1.1 \\ 
145 & $-$46$\pm$\ \ 3 &  82$\pm$\ \ 3 & 10700$\pm$520 &  3.54$\pm$0.12 & 9660$\pm$\ \ 60 &  3.42$\pm$0.03 & $-$1.0 \\ 
147 & $-$54$\pm$\ \ 4 &  73$\pm$\ \ 5 & 10800$\pm$550 &  3.71$\pm$0.12 & 9810$\pm$\ \ 80 &  3.43$\pm$0.04 & $-$0.8 \\ 
149 & $-$67$\pm$\ \ 5 &  78$\pm$\ \ 5 & 11100$\pm$590 &  3.58$\pm$0.17 & 9800$\pm$\ \ 80 &  3.41$\pm$0.04 & $-$1.0 \\ 
151 & $-$53$\pm$\ \ 4 &  72$\pm$\ \ 4 & 10800$\pm$540 &  3.74$\pm$0.11 & 9760$\pm$\ \ 80 &  3.43$\pm$0.04 & $-$0.9 \\ 
154 & $-$51$\pm$\ \ 3 &  88$\pm$\ \ 4 & 10800$\pm$520 &  3.39$\pm$0.15 & 10100$\pm$100 &  3.51$\pm$0.04 & $-$1.1 \\ 
156 & $-$46$\pm$\ \ 8 &  88$\pm$10 & 10700$\pm$610 &  3.40$\pm$0.22 & 9760$\pm$\ \ 80 &  3.37$\pm$0.04 & $-$1.0 \\ 
157 & $-$52$\pm$11    &  79$\pm$13 & 10800$\pm$670 &  3.59$\pm$0.23 & 9580$\pm$\ \ 90 &  3.34$\pm$0.05 & $-$1.1 \\ 
169 & $-$88$\pm$\ \ 4 &  72$\pm$\ \ 5 & 11600$\pm$600 &  3.69$\pm$0.17 & 10400$\pm$100 &  3.53$\pm$0.05 & $-$1.0 \\ 
176 & $-$72$\pm$\ \ 5 &  80$\pm$\ \ 6 & 11200$\pm$600 &  3.52$\pm$0.19 & 10300$\pm$100 &  3.52$\pm$0.05 & $-$1.0 \\ 
180 & $-$75$\pm$\ \ 4 &  82$\pm$\ \ 5 & 11300$\pm$580 &  3.46$\pm$0.18 & 10600$\pm$\ \ 80 &  3.60$\pm$0.04 & $-$1.1 \\ 
\hline
\hline
\end{tabular}
\end{table*}

\begin{figure}[h!]
\includegraphics[height=\columnwidth,angle=270]{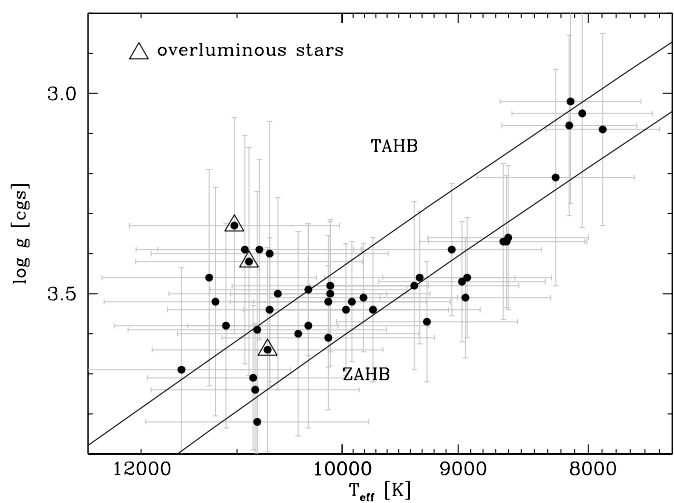}
\caption[]{Atmospheric parameters derived from the $a_0, sz$ indices
  for stars that do not show evidence for radiative levitation. For
  comparison we also show curves representing a canonical zero-age
  horizontal branch and terminal-age horizontal branch from \citet{mola03}.
\label{fig:Teff_g_phot} }
\end{figure}

To restrict the fitting range we used only $\log g$ between 3
and 4 and $T_{\rm eff}$ between 8\,000\,K and 12\,000\,K. We fitted
a second-order polynomial to the relation $T_{\rm eff}\,{\rm (a_0)}$, which yielded an rms deviation of 30\,K. For the
surface gravities we fitted second-order polynomials to the
relation $\log g\,{\rm (a_0, sz)}$, which yielded an rms deviation of
0.04 in $\log g$. Temperatures and surface gravities derived
from these relations are listed in Table\,\ref{tab:par_phot} and
plotted in Fig.\,\ref{fig:Teff_g_phot} for stars cooler than the
$u$-jump.

The errors provided in Table\,\ref{tab:par_phot} are a quadratic
combination of errors from the photometric data, the uncertainty of
the reddening in the $y$-band (estimated to be 0\fm02) and the
error of the fit to the theoretical relations. To  
estimate the influence of metallicity we compared the results
obtained for the two metallicities mentioned above. The differences in
temperature vary almost linearly from $-$0.1\% at 8000\,K to $+$0.8\%
at 12\,000\,K, with the results from the more metal-poor models being
hotter at the hot end. The differences are therefore well below the errors
on the individual temperatures.

The differences in surface gravities for these two metallicities show
a quadratic dependency on the effective temperature, varying from
  $+$0.02 at about 8\,600\,K (with the surface gravities from the
  metal-poor models being higher at 8\,600\,K than those from the more
  metal-rich models) to 0 at 9\,000\,K and 11\,000\,K via $-$0.02 at
  about 10\,000\,K.  For temperatures below 8\,500\,K the differences
  increase to $-$0.08, but these stars are too cool to be analysed
  spectroscopically with our methods (see
  Sect.\,\ref{sec:lineprof}). Within the temperature range of interest
  for further analysis the differences are below the average errors
of the surface gravities given in Table\,\ref{tab:par_phot}. For
  reasons of consistency between photometric and spectroscopic
  analysis we decided against averaging the photometric grids.

From the errors listed in Table\,\ref{tab:par_phot} and the
uncertainties due to the metallicity, we estimate typical errors in
$T_{\rm eff}$ and $\log g$ to be about 5\% and 0.13 dex, respectively.

\subsection{Line profile fitting with homogeneous model spectra} \label{sec:lineprof}

For stars redder than the $u$-jump, we computed model atmospheres with
convection for [M/H] = $-$1.5 using ATLAS9 \citep{kuru93} and used
\href{http://www.sternwarte.uni-erlangen.de/~ai26/linfit/linfor.html}{Lemke's
  version}
of the LINFOR program (developed originally by Holweger, Steffen, and
Steenbock at Kiel University) to compute a grid of theoretical spectra
that include the Balmer lines H\,$\alpha$ to H$_{22}$ and the
\ion{He}{i} and \ion{He}{ii} lines. The grid covers the range
7\,500\,K~$\leq T_{\rm eff} \leq$~12\,000\,K, $2.5~\leq \log g
\leq5.0$, and $-3.0\leq {\log{\frac{n_{\rm He}}{n_{\rm H}}}}
\leq-1.0$. For stars bluer than the $u$-jump we used model atmospheres
computed for [M/H] = +0.5 (see \citealt{mosw00} for details).  From an
extrapolation of the LTE/NLTE threshold for subdwarf B stars
\citep{napi97} we assumed that LTE is a valid approximation here as
well. To establish the best fit to the observed spectra, we used the
fitsb2 routines developed by \citet{nayu04}, which employ a $\chi^2$
test. The $\sigma$ necessary to calculate $\chi^2$ was estimated from
the noise in the continuum regions of the spectra. The fit program
normalises synthetic model spectra {\em and} observed spectra using
the same points for the continuum definition.

While the errors were obtained via boot strapping and should therefore
be rather realistic, they do not include possible systematic errors
due to flat-field inaccuracies or imperfect sky subtraction, for
instance. The true errors in $T_{\rm eff}$ are probably close to those
from photometry, that is, 5\%, and the true errors in $\log g$ are
probably about 0.1.

Because the $\chi^{2}$ fit of the line profile can lead to ambiguous
results for effective temperatures close to the Balmer maximum, the
results from Sect.\,\ref{sec:phot_par} were taken as initial
parameters for the spectral line profile fitting procedure.

In Table\,\ref{tab:par_phot} we list the results obtained from
fitting the Balmer lines H$\beta$ to H$_{10}$ (excluding H$\epsilon$
to avoid the \ion{Ca}{ii}~H line) in the cool stars. For the cool
stars we did not fit the helium lines because they are very weak and tend
to produce spurious results. We verified, however, that the helium
lines predicted by the model spectra with solar helium abundance did
reproduce the observations. The reddest stars that showed many strong
metal lines (55, 60, 61, 63, 72) were omitted from the line profile
fits because our model spectra contain only H and He lines.  In addition, we fitted the \ion{He}{i}
lines at $\lambda\lambda$ 4026\AA, 4388\AA, 4472\AA, and 4922\AA\
for stars hotter than 11\,500\,K.

\begin{table}
\caption[]{Temperatures, surface gravities, and helium abundances from
  line profile fits for stars bluer than the $u$-jump that show
  evidence of radiative levitation. The errors are derived from the
  $\chi^2$ fit (see Sect.\,\ref{sec:lineprof} for details) and are
  only statistical errors. In addition, we list the helium abundance as
  determined with LINFOR (see Sect.\,\ref{sec:abu} for details).
\label{tab:par_spec_hot}}
\begin{tabular}{rrrrr}
\hline
\hline
ID & $T_{\rm eff}$ & $\log g$ 
&  \multicolumn{2}{c}{$\log{({\frac{n_{\rm He}}{n_{\rm H}}})}$} \\
 & [K] & & fitsb2 & LINFOR\\ 
\hline
100 & 11400$\pm$100 &  3.65$\pm$0.03  & $-$2.23$\pm$0.11 & $-$2.14 \\ 
101 & 11400$\pm$100 &  3.78$\pm$0.03  & $-$1.95$\pm$0.11 & $-$1.93 \\ 
142 & 11400$\pm$100 &  3.72$\pm$0.03  & $-$2.98$\pm$0.12 & $-$2.76 \\ 
146 & 11600$\pm$100 &  3.81$\pm$0.03  & $-$2.06$\pm$0.10 & $-$2.02 \\ 
179 & 11600$\pm$100 &  3.77$\pm$0.03  & $-$2.37$\pm$0.10 & $-$2.14 \\ 
183 & 12300$\pm$100 &  3.93$\pm$0.03  & $-$2.27$\pm$0.12 & $-$2.36 \\ 
187 & 11700$\pm$100 &  3.84$\pm$0.03  & $-$2.74$\pm$0.13 & $-$2.87 \\ 
195 & 12400$\pm$100 &  3.86$\pm$0.03  & $-$2.55$\pm$0.15 & $-$2.51 \\ 
196 & 12300$\pm$100 &  3.91$\pm$0.03  & $-$2.67$\pm$0.18 & $-$2.49 \\ 
199 & 12400$\pm$100 &  3.94$\pm$0.03  & $-$2.79$\pm$0.18 & $-$2.84 \\ 
212 & 12900$\pm$100 &  3.93$\pm$0.02  & $-$2.66$\pm$0.12 & $-$2.57 \\ 
213 & 13100$\pm$100 &  3.97$\pm$0.03  & $-$2.61$\pm$0.15 & $-$2.68 \\ 
216 & 12900$\pm$100 &  3.94$\pm$0.02  & $-$2.66$\pm$0.14 & $-$2.71 \\ 
221 & 13200$\pm$100 &  3.97$\pm$0.03  & $-$2.67$\pm$0.13 & $-$2.60 \\ 
228 & 13800$\pm$100 &  4.00$\pm$0.03  & $-$2.43$\pm$0.17 & $-$2.43 \\ 
230 & 13800$\pm$100 &  4.06$\pm$0.03  & $-$2.28$\pm$0.10 & $-$2.34 \\ 
231 & 13300$\pm$100 &  4.00$\pm$0.02  & $-$2.72$\pm$0.09 & $-$2.55 \\ 
240 & 13300$\pm$100 &  3.99$\pm$0.03  & $-$2.71$\pm$0.11 & $-$2.61 \\ 
242 & 14300$\pm$100 &  4.03$\pm$0.03  & $-$2.37$\pm$0.13 & $-$2.28 \\ 
243 & 14400$\pm$100 &  4.09$\pm$0.04  & $-$2.52$\pm$0.15 & $-$2.43 \\ 
275 & 15400$\pm$200 &  4.16$\pm$0.03  & $-$2.44$\pm$0.15 & $-$2.27 \\ 
288 & 15200$\pm$200 &  4.14$\pm$0.04  & $-$2.19$\pm$0.11 & $-$2.16 \\ 
292 & 15100$\pm$100 &  4.23$\pm$0.03  & $-$2.92$\pm$0.07 & $-$2.80 \\ 
300 & 15400$\pm$200 &  4.14$\pm$0.04  & $-$2.21$\pm$0.10 & $-$2.19 \\ 
304 & 16900$\pm$200 &  4.29$\pm$0.04  & $-$2.57$\pm$0.12 & $-$2.46 \\ 
347 & 16400$\pm$200 &  4.38$\pm$0.04  & $-$2.11$\pm$0.09 & $-$2.06 \\ 
\hline
\end{tabular}
\end{table}

\begin{figure}[h!]
\includegraphics[height=\columnwidth,angle=270]{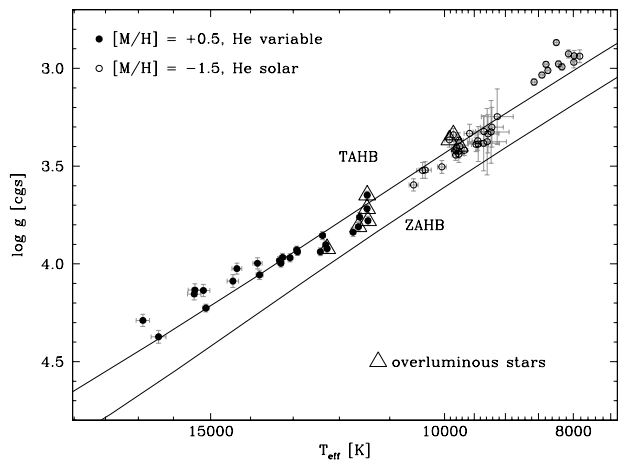}
\caption[]{Atmospheric parameters derived from the line profile
  fitting for stars hotter (full symbols) and cooler (open symbols)
  than the $u$-jump.  For comparison we also show a canonical zero-age
  horizontal branch and terminal-age horizontal branch from
  \citet{mola03}. \label{fig:Teff_g_hot} }
\end{figure}

Figure\,\ref{fig:Teff_g_hot} shows a gap at 8\,500\,K\,$\le T_{\rm
  eff}\le$\,9\,000\,K in the atmospheric parameters derived from line
profile fitting, which was first noted and discussed by \citet{mola03}
for observations of HB stars in M\,13.  Stars populating this region
in Fig.\,\ref{fig:Teff_g_phot} close to the zero-age horizontal branch
({\bf ZAHB}) are located at about 8\,000\,K in
Fig.\,\ref{fig:Teff_g_hot}, above the terminal-age horizontal branch
({\bf TAHB}). Prompted by the referee, we treated a homogeneous model
spectrum calculated for $T_{\rm eff}$ = 8\,800\,K,$\log g$ = 3.35 and
[M/H] = $-$1.5 as described in Sect.\,\ref{ssec:strat} to simulate an
observed spectrum. Varying the starting values of the fit from
8\,000\,K to 9\,500\,K for $T_{\rm eff}$ and 3.2 to 3.6 for $\log g$
did not affect the fit results, which remained at $T_{\rm eff}$ =
8\,841\,K and $\log g$ = 3.35. Varying the FWHM of the Gaussian with
which the model spectra are convolved before fitting from 0.6\,\AA\ to
0.75\,\AA\ resulted in $T_{\rm eff}$ between 8\,818\,K and 8\,862\,K
and $\log g$ between 3.34 and 3.36. We noted, however, that the
resolution of the observed data seems to vary slightly with wavelength
--- at the blue end the predicted line profiles are very slightly
wider and deeper than the observed ones, while for H$\beta$ the
observed line core is slightly more narrow and deep. To rule out
possible small variations in resolution across the wavelength range as
the cause for the gap, we convolved the observed spectra to a
resolution of 2.5\,\AA, which is close to the resolution of the data
of \citet{momo07,momo09,movi11} and \citet{samo13}, which do not show
such a gap in their results, although they used the same analysis
methods as we did. Unfortunately, the gap remains in the results of
the line profile fitting. Using model atmospheres with or without
convection did not affect the gap either.

 The stars in Fig.\,\ref{fig:Teff_g_hot} can be split into two groups
 with respect to their position relative to the evolutionary
 sequences: stars with effective temperatures between 9\,000\,K and
 roughly 14\,000\,K lie between the ZAHB and the TAHB, while stars
 hotter or cooler than this temperature range lie above the
 TAHB.

\begin{figure}[h!]
\includegraphics[width=\columnwidth,angle=0]{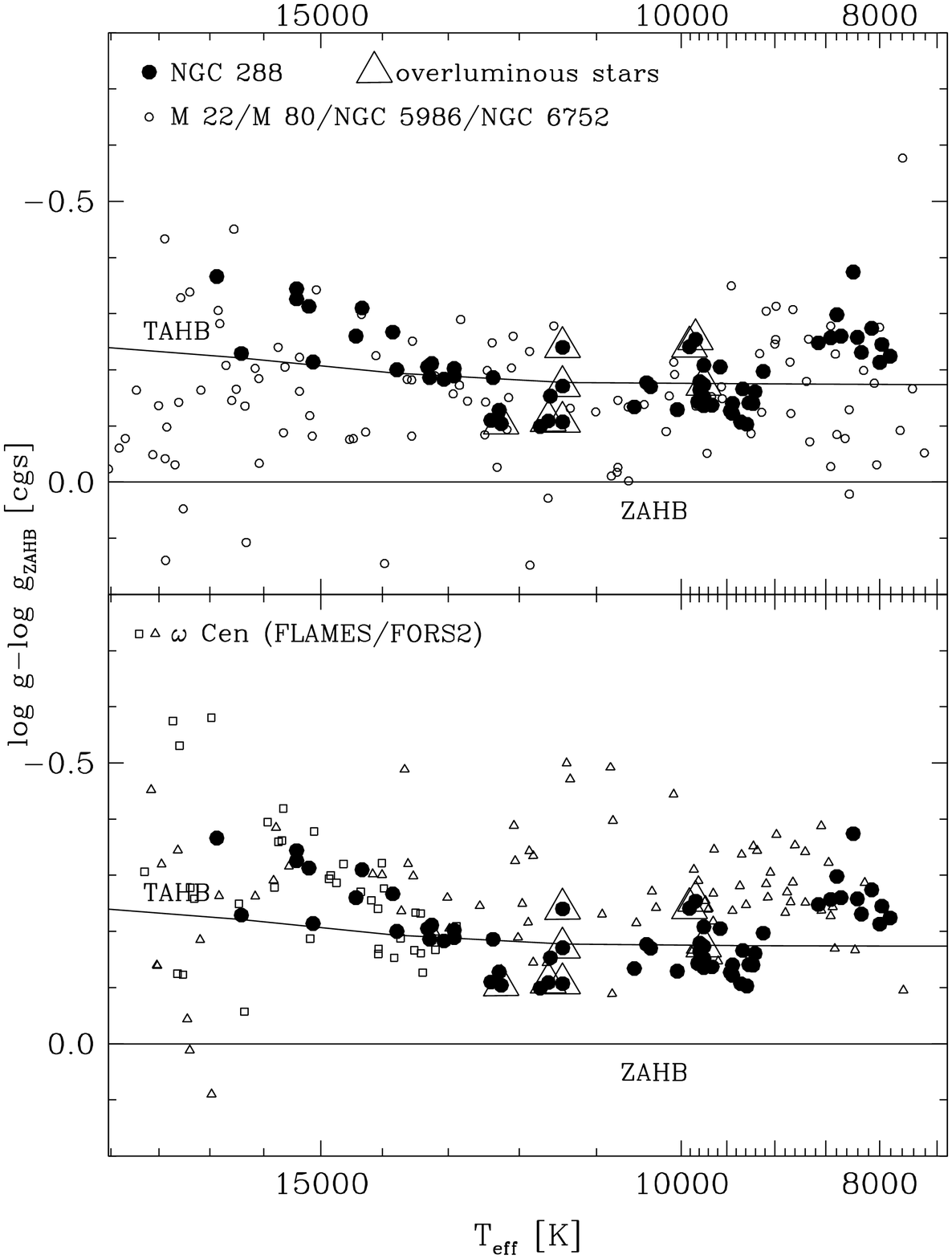}
\caption[]{Surface gravity relative to the zero-age horizontal branch
  at the given effective temperature. For comparison we also show a
  canonical zero-age and terminal-age horizontal branch from
  \citet{mola03}. In the upper plot we also show results from FORS2
  observations of hot HB stars in M\,80, NGC\,5986 \citep{momo09},
  NGC\,6752 \citep{momo07}, and M\,22 \citep{samo13}.  In the lower
  plot we also provide the results obtained from FLAMES and FORS2
  observations of hot horizontal branch stars in $\omega$\,Cen
  (\citealt{modr11}, small squares; \citealt{movi11}, small
  triangles).\label{fig:Teff_difflogg}}
\end{figure}

To determine how the parameters derived for the hot HB stars in
NGC\,288 differ from those obtained for other clusters and from the
theoretical predictions we calculated the difference between the
derived surface gravity and that on the ZAHB for the given
temperature. Figure\,\ref{fig:Teff_difflogg} compares our results with
those obtained for M\,80, NGC\,5986 \citep{momo09}, NGC\,6752
\citep{momo07}, M\,22 \citep{samo13}, and $\omega$\,Cen
\citep{modr11,movi11}. Here the NGC\,288 results resemble those from
$\omega$\,Cen for temperatures lower than 9\,000\,K or higher than
about 14\,000\,K, while they coincide with the other clusters for
temperatures between 9\,000\,K and 14\,000\,K.

\begin{figure}[h!]
\includegraphics[width=\columnwidth,angle=0]{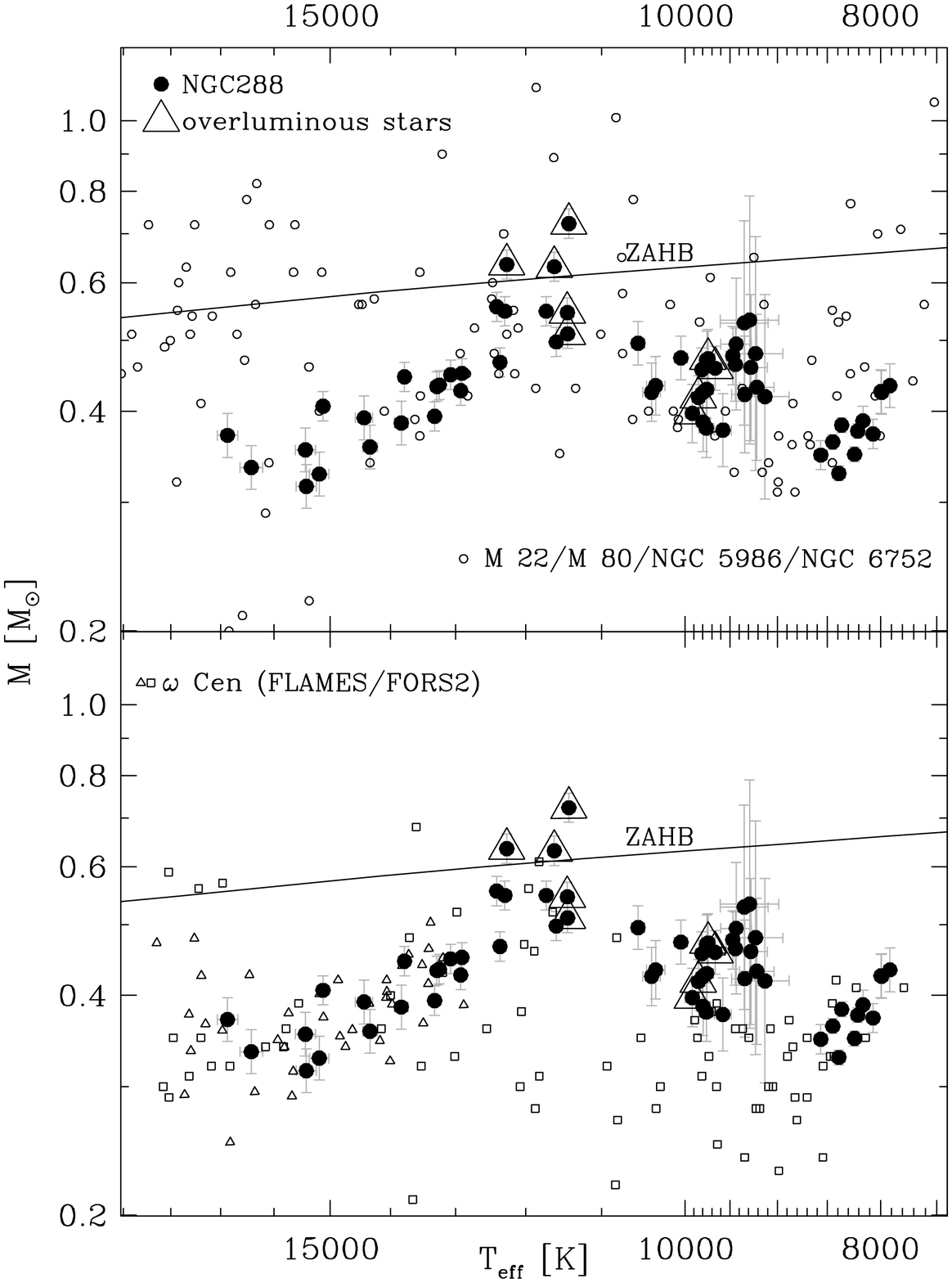}
\caption[]{Masses from line profile fits.  For comparison we also show
  a canonical zero-age horizontal branch from \citet{mola03}. In the
  upper plot we also show results from FORS2 observations of hot HB
  stars in M\,80, NGC\,5986 \citep{momo09}, NGC\,6752 \citep{momo07},
  and M\,22 \citep{samo13}.  In the lower plot we also provide the
  results obtained from FLAMES and FORS2 observations of hot
  horizontal branch stars in $\omega$\,Cen (\citealt{modr11}, small
  squares; \citealt{movi11}, small triangles).\label{fig:Teff_mass}}
\end{figure}

Using the spectroscopically determined atmospheric parameters, $y$
-magnitudes, and the distance modulus of the globular cluster
($(m-M){\rm _V}$\,=\,14\fm95, \citealt{cagr00}), we derived the masses shown in
Fig.\,\ref{fig:Teff_mass} using the equation
\begin{equation}
{\frac{M}{M_\sun}} = {\frac{3.6\cdot10^{-7}}{\pi \cdot
    g_\sun \cdot R_\sun
  [{\rm pc}]}} \cdot 10^{-0.4 \cdot (y-(m-M)_V-y_{\rm th}),} 
\end{equation}
where $y_{\rm th}$ is the theoretical $y$ magnitude at the stellar
surface from \citet{kuru93} and $R_\sun$ is the solar radius in
parsec.  Obviously, the masses are systematically too low, except for
stars just hotter than the $u$-jump. Compared with results from other
clusters, the masses of the stars in NGC\,288 show a similar behaviour
as the surface gravities with respect to HB stars in other globular
clusters.

\section{Abundances}\label{sec:abu}
We used the parameters derived in Sect.\,\ref{sec:lineprof} for
  the following abundance analysis.
\subsection{Helium}\label{ssec:helium}
Helium abundances were already derived during the determination of
effective temperatures and surface gravities. In addition, we
determined them (together with abundances for other elements) via
spectrum synthesis using the abundance-fitting routine of LINFOR (see
Sect.\,\ref{ssec:metals} for details). The resulting abundances are
listed in Tables\,\ref{tab:par_phot} and \ref{tab:par_spec_hot}. For
the cool stars the abundances should be treated with caution because
the helium lines are rather weak. However, the average fitted helium
abundance ${\rm \log{\frac{n_{He}}{n_{H}}}}$ of $-$1.03 agrees well
with our assumption of solar helium abundance for cool stars. For hot
stars the helium abundances derived with LINFOR are in general
slightly higher than those derived with fitsb2, but in 80\% of the
cases the difference is smaller than the error provided by
fitsb2. On average, the LINFOR abundances are higher by 0.06\,dex,
that is, much smaller than the error of the fitsb2 results.

\begin{figure}[!h]
\includegraphics[height=\columnwidth,angle=270]{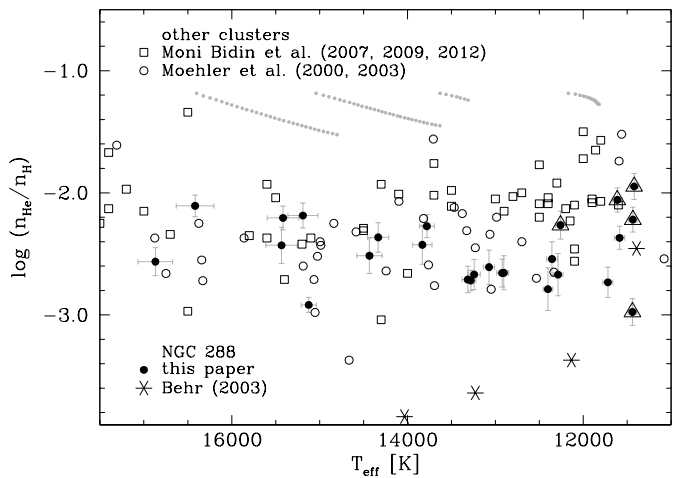}
\caption[]{Helium abundances for hot HB stars in M\,80, NGC\,5986,
  NGC\,6752, and $\omega$\,Cen (Moni Bidin et al.), M\,3, M\,13, and
  NGC\,6752 (Moehler et al.) and NGC\,288.  The grey dots mark the
  abundances predicted for various HB ages from diffusion theory (with
  an ad hoc surface-mixing zone, \citealt{miri11}, see
  Sect.\,\ref{ssec:diff} for details). The tracks have dots every 3
  Myr.}\label{Fig:helium}
\end{figure}

Because helium abundances have been derived for HB stars in many
globular clusters with similar methods and model spectra as we used
here, we compared our results with published data in
Fig.\,\ref{Fig:helium}. The helium abundances of
\citet{momo07,momo09,movi12} and \citet{mosw00,mola03} were derived
from low-resolution data with a resolution of 2.6\,\AA --
3.4\,\AA. The abundances of \citet{behr03} were derived from
data with a resolution of 0.1\,\AA, while the data used here have a
resolution of about 0.7\,\AA. Between 11\,000\,K and 13\,500\,K the
abundances show a clear correlation with the resolution of the data
from which they were derived -- the abundances decrease with
increasing resolving power. This behaviour had been noticed already by
\citet{movi12}. This may be caused by the fact that
at these relatively cool temperatures the helium lines are rather
weak, which together with the helium deficiency makes them hard to
fit. For higher temperatures the abundances from the different sources
overlap, with the exception of the hottest star of
\citet{behr03}. Due to the small number of results from
high-resolution spectroscopy it is not clear whether the resolution
effect persists to higher temperatures. For a comparison with
predictions from diffusion theory see Sect.\,\ref{ssec:diff}.

\subsection{Heavy elements}\label{ssec:metals}
\begin{figure}[h!]
\includegraphics[width=\columnwidth,angle=0]{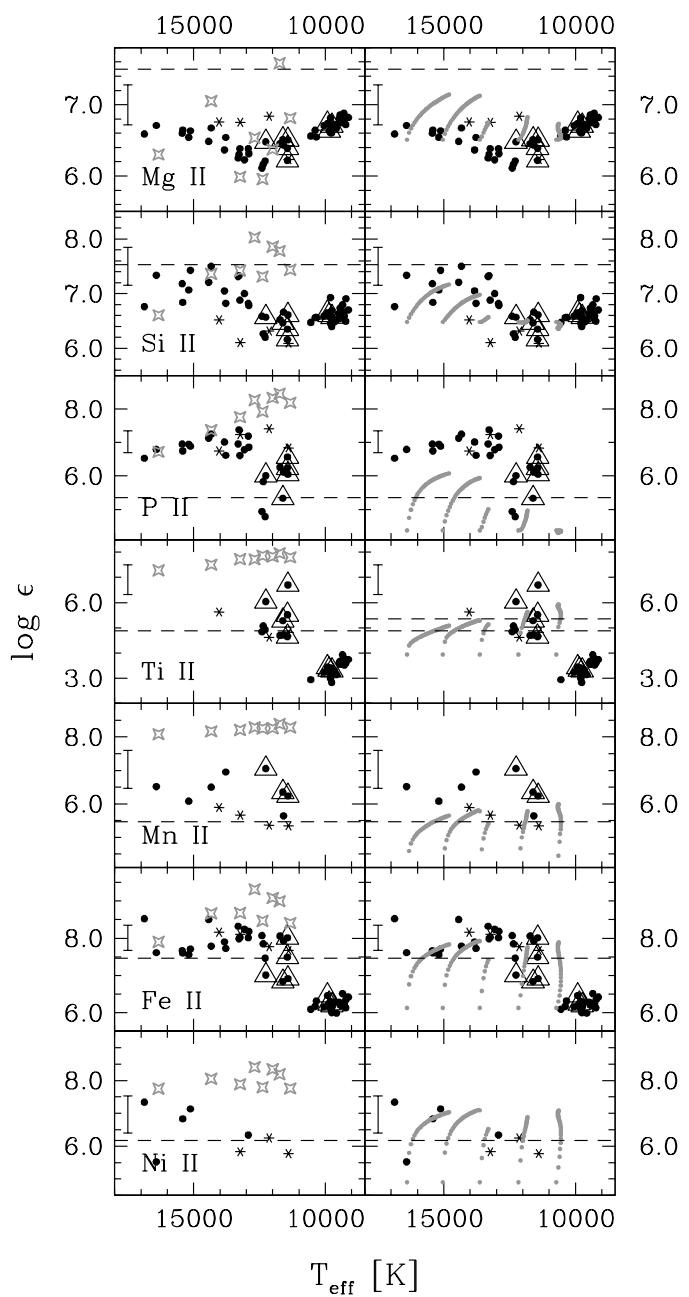}
\caption[]{\color{black}Abundances derived via spectrum synthesis for
  all stars hotter than 9\,000\,K. The triangles mark the same stars
  as in Fig.\,\ref{fig:cmd_hb}. The asterisks mark the results from
  \citet{behr03}. The bars at 17\,500\,K mark the average error bars
  of \citet{behr03} for NGC\,288. Our errors are probably not
  smaller. The dashed lines mark the solar abundances. The
  four-pointed stars mark the abundances derived for stratified model
  spectra (left column, see Sect.\,\ref{ssec:strat} for details), which
  indicate the equilibrium abundances achievable by diffusion. The
  grey dots (right column) mark the abundances predicted for various
  HB ages from diffusion theory (with an ad hoc surface-mixing zone,
  \citealt{miri11}, see Sect.\,\ref{ssec:diff} for
  details). The tracks have dots every 1 Myr.}\label{Fig:abu}
\end{figure}

Although the spectra {\bf have only} medium resolution, we estimated
abundances via spectrum synthesis using the abundance-fitting routine
of LINFOR to verify whether our assumptions about the overall
increase in heavy elements were roughly correct. We used the line lists
from \citet{kuru93} and simultaneously determined abundances
for He\,{\sc i}, Mg\,{\sc ii}, Si\,{\sc ii}, P\,{\sc ii}, Ti\,{\sc
  ii}, Mn\,{\sc ii}, Fe\,{\sc ii}, and Ni\,{\sc ii}. When the fitted
abundances did not change anymore from one iteration to the next, we
visually verified that the spectrum had indeed been
well reproduced. The results are shown in Fig.\,\ref{Fig:abu}. 
  The LINFOR abundance-fitting routine only provides an overall error,
  but the scatter of abundances below 11\,000\,K gives an idea of the
  minimum uncertainties, because we do not expect the abundances shown
  in Fig.\,\ref{Fig:abu} to vary in {\bf these} cool stars. For stars
  hotter than the $u$-jump additional uncertainties arise because we used homogeneous model atmospheres to analyse stars affected
  by diffusion.
                                 
The results for P\,{\sc ii}, Ti\,{\sc ii}, Mn\,{\sc ii}, Fe\,{\sc ii}, and
Ni\,{\sc ii} approximately agree with our assumption of [M/H] = $+$0.5 
  for stars hotter than the $u$-jump, while
Mg\,{\sc ii} and Si\,{\sc ii}, on the other hand, are rather lower than
our assumed abundances. However, since the atmospheric structure
depends more on iron than on the light elements, we took these results
to indicate that our assumptions were reasonable. For the cool
  stars, Mg\,{\sc ii} and Ti\,{\sc ii} seem to show an unexpected gradient with
  temperature, but it is unclear whether this is significant.

Our results show some differences compared with those obtained for
  NGC\,288 by \citet{behr03}. \citet{behr03}
  found no clear dependence of the Mg abundance relative to $T_{\rm
    eff}$ for the three clusters studied with a sufficient number of
  stars with $T_{\rm eff}$ above the $u$-jump, while a trend appears
  to be present in our results for NGC\,288. For Si and Mn Behr found
  significantly lower abundances than we do, which might be an effect
  of instrumental resolution (see Sect.\,\ref{ssec:helium}). 
The observed abundances shown in Fig.\,\ref{Fig:abu} are provided
  at the CDS.
\subsection{Abundances predicted by stratified model spectra}\label{ssec:strat}
\begin{figure}[h!]
\centering
\includegraphics[width=\columnwidth,angle=00]{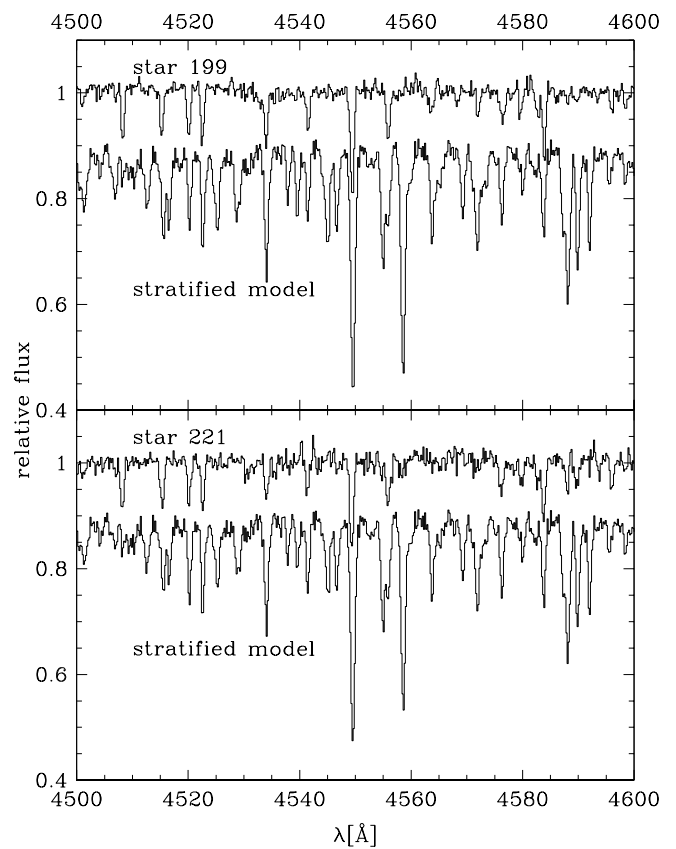}
\caption[]{Normalised spectra for stars 199 (upper panel, 12\,400\,K)
  and 221 (lower panel, 13\,400\,K) compared with stratified model
  spectra F (upper panel) and H (lower panel), which have fitted
  parameters close to those of the observed spectra. For clarity,
  the stratified model spectra have been offset by 0.1 from the
  observed spectra (see Tables\,\ref{tab:par_spec_hot} and
  \ref{tab:strat_homo}).\label{Fig:obs_strat}}\end{figure}

\color{black} To compare the results from our observed spectra with
the predictions from stratification theory we also determined
abundances from model spectra of \citet{lmhh09}, which include
abundance stratification due to diffusion. These models
self-consistently calculate the structure of the atmosphere with the
vertical abundance stratification. The abundances of the individual
elements are calculated in each layer of the atmosphere while assuming
equilibrium (nil diffusion-velocity). This leads to vertical abundance
stratification and modifies the atmospheric structure.  First we
compared the observed metal lines with those predicted by 
  stratification models and found that the observed lines were much
weaker than predicted by these models (see Fig.\,\ref{Fig:obs_strat}).

As discussed in \citet{lmhh09}, the abundances of some elements (Fe,
for example) predicted by the models can be overestimated because in a
real star the diffusion process is a time-dependent phenomenon.  Even
though the radiative forces on a given element can theoretically
support a very large overabundance, this situation may be hampered by
various factors. The equilibrium solution used in these models may
therefore not be reached. For instance, in a real star, such a large
quantity of atoms might not be able to surface due to the diffusion
that takes place below the stellar atmosphere.  In addition, the
abundances of other elements (such as helium) that have relatively
weak radiative accelerations can be underestimated by these models.

Next we convolved the stratified model spectra to the same resolution
as our observed data and multiplied them with a spectrum of average 1
and rms of 0.0125, resulting in a signal-to-noise ratio of 80. Then we
determined their effective temperatures and surface gravities in the
same way as described in Sect.\,\ref{sec:lineprof}, except that we
used a model grid with [M/H] = $+$1, since the lines in the stratified
model spectra were much stronger than in the observed spectra (see
Fig.\,\ref{Fig:obs_strat}). The resulting parameters are given in
Table\,\ref{tab:strat_homo} and clearly show that the analysis of
  stratified model spectra with homogeneous model spectra yields lower
  temperatures and surface gravities, with the difference in
  temperature increasing with increasing temperature.

\begin{table}
\caption[]{Temperatures and surface gravities obtained from
    fitting stratified model spectra with homogeneous model spectra
    for [M/H] =$+$1 (see Sect.\,\ref{ssec:strat} for details). The
  errors are derived from the $\chi^2$ fit (see
  Sect.\,\ref{sec:lineprof} for details) and are only statistical errors.
\label{tab:strat_homo}}
\begin{tabular}{crr|rr}
\hline
\hline
\multicolumn{3}{c|}{model} & \multicolumn{2}{c}{fit}\\
 &  $T_{\rm eff}$ & $\log g$& $T_{\rm eff}$ & $\log g$\\
 & [K] & & [K] & \\
\hline
A & 12000 & 3.5 & 11500$\pm$100 & 3.07$\pm$0.05\\
B & 12000 & 4.0 & 11700$\pm$100 & 3.60$\pm$0.06\\
C & 12000 & 4.5 & 11300$\pm$100 & 3.94$\pm$0.06\\
D & 13000 & 3.5 & 12200$\pm$100 & 3.08$\pm$0.05\\
E & 13000 & 4.0 & 12000$\pm$100 & 3.48$\pm$0.05\\
F & 13000 & 4.5 & 12400$\pm$100 & 4.00$\pm$0.05\\
G & 14000 & 4.0 & 12700$\pm$100 & 3.47$\pm$0.04\\
H & 14000 & 4.5 & 13200$\pm$100 & 4.01$\pm$0.04\\
I & 16000 & 4.5 & 14300$\pm$100 & 3.98$\pm$0.04\\
J & 18000 & 5.0 & 16300$\pm$100 & 4.52$\pm$0.03\\ 
\hline
\end{tabular}
\end{table}

Using the parameters in the right column of
Table\,\ref{tab:strat_homo}, we then determined abundances for the
model spectra with fitted $\log$\,g above 3.45 as expected for stars
on or near the ZAHB at the fitted temperatures, that is, all models
except A and D. We alternated between simultaneously fitting the ions
Mg\,{\sc ii}, Si\,{\sc ii}, P\,{\sc ii}, S\,{\sc ii}, A\,{\sc ii},
Ti\,{\sc ii}, V\,{\sc ii}, Cr\,{\sc ii}, and simultaneously fitting
the ions Mn\,{\sc ii}, Fe\,{\sc i}, Fe\,{\sc ii}, Co\,{\sc ii},
Ni\,{\sc ii}, Sr\,{\sc ii}, and Zr\,{\sc ii}, until the fit did not
improve any more. Then we again visually checked that most of the
spectrum had been reproduced. Typically, we found that some lines were
not fit well, but that was to be expected since the abundance
distribution in the model atmospheres creating these spectra is far
from homogeneous, which results in a very different atmospheric
structure. The results of these fits are marked by four-pointed stars
in the left column of Fig.\,\ref{Fig:abu}. As already suggested by the
comparison shown in Fig.\,\ref{Fig:obs_strat}, the abundances derived
from the stratified spectra are generally higher than those derived
from the observed spectra, with He\,{\sc i} (virtually absent from
stratified model atmospheres) and Mg\,{\sc ii} being the exceptions.

\subsection{Abundances predicted by diffusion models}\label{ssec:diff}

It is also possible to compare our abundance results with stellar
evolution models computed including the effects of atomic diffusion
\citep{miri08,miri11}. These models followed the evolution from the
zero-age main sequence, treating in detail atomic diffusion inside the
star. On the horizontal branch, they predicted large overabundances of
metals, often larger than observed above 11\,000\,K (similar to the
problems found above for the stratified model atmospheres). A surface
mixed mass of around 10$^{-7}\,M_\sun$ was needed to reduce the
expected anomalies to values observed in HB stars of a number of
clusters as well as in sdB stars. However, the introduction of a
surface mixing layer leads to a vertically homogeneous
atmosphere. This contradicts the observed vertical stratification of
certain elements, including iron, in some blue HB stars
\citep{klbw07,klbw08,khlb10}.  The tracks shown in Fig.\,\ref{Fig:abu}
are taken from \citet{miri11}. These tracks start with the original
cluster abundances at the ZAHB and cover the first third of horizontal
branch evolution, which lasts some 100 Myr\footnote{The models around
  12\,000 and 13\,500\,K were stopped around 10\,Myr because of
  convergence problems, but the trend can be estimated from the
  surrounding models.}.

Figure\,\ref{Fig:abu} shows that the abundances change more rapidly
during the first 10\,Myr than during the following 20\,Myr for all
species that become overabundant. This is largely caused by a
reduction of the radiative accelerations as the concentrations
increase.  The abundance increase is also expected to be slow during
the following 70\,Myr for these species. The Si, Ti, Fe, and Ni
  observations are within the expected abundance range when one takes
  error bars evaluated from the scatter of Fe and Si below 11\,000\,K
  into account. For Si, the trend suggested by the models
    appears to be present in the data. For Mn our abundances are
  higher than the results from \citet{behr03}, which
  agree well with the predictions. Phosphorus is some five times more
overabundant than predicted by the models, whereas magnesium shows
small effects, while the models predict about five times larger abundances.

Helium is observed to be more underabundant than expected after
30\,Myr (cf. Fig.\,\ref{Fig:helium}).  However, its abundance has a
very different time dependence from that of metals.  The
underabundance of helium is caused by gravitational settling, and this
process is as rapid from 10 to 30\,Myr as during the first
10\,Myr. Because helium settles, the settling goes as e$^{-t/\theta}$
, where $\theta$ is the settling time scale.  This means that the
settling continues exponentially, and if helium is underabundant by a
factor of 3 after 30\,Myr it should be underabundant by a factor of
around 27 after 100\,Myr, corresponding to a value of 9.56 in
Fig.\,\ref{Fig:abu} at the end of the horizontal branch
evolution. Because it is highly unlikely, however, that most of the
stars in Fig.\,\ref{Fig:abu} are at the end of their HB evolution,
some discrepancy remains.

As may be seen from Fig.\,4 of \citet{miri11}, abundance anomalies
caused by atomic diffusion are not limited to a thin surface
phenomenon.  However, there are species (e.g. Mg) whose observed
anomalies do not agree with the expected ones, suggesting that the
model may be missing something. Assuming that the outer
10$^{-7}\,M_\sun$ is completely mixed could be an oversimplification,
especially since the mixing mechanism is currently unknown. For
instance, it might be thermohaline convection (see Sect 5.3 of 
\citealt{miri11}).  This is a relatively weak convection that might
not eliminate all effects of additional diffusion in the atmosphere.
For instance, helium is largely neutral in the atmosphere of HB stars
of 11\,000\,K to 15\,000\,K so that it has an atomic diffusion
coefficient larger than that of ionized metals (by a factor of around
100); helium might then be the most affected atomic species as
turbulence weakens.  This remains speculative.

\section{Line profile fitting with stratified model spectra} \label{sec:lineprofstrat}
\begin{figure}[h!]
\includegraphics[height=\columnwidth,angle=270]{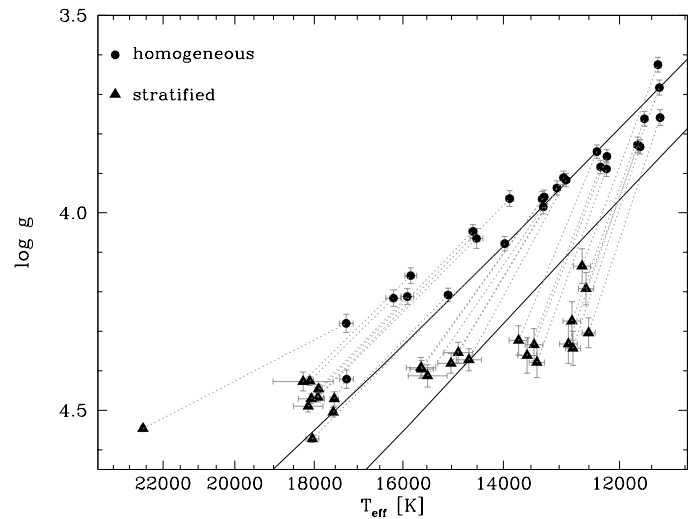}
\caption[]{Atmospheric parameters derived from the line profile
  fitting for stars hotter than the $u$-jump, using homogeneous model
  atmospheres ([M/H] = $+$0.5, ${\log{\frac{n_{\rm He}}{n_{\rm H}}}}$
  = $-$3, filled circles) and stratified model atmospheres (filled
  triangles).  For comparison we also show a canonical zero-age
  horizontal branch and terminal-age horizontal branch from
  \citet{mola03}. \label{fig:Teff_g_strat} }
\end{figure}

 We also examined how the use of stratified instead of homogeneous
 model atmospheres to fit our observations influenced the parameters
 determined from these fits.  To allow a direct comparison with
 results obtained using homogeneous model spectra without metal lines
 to fit our observations, we calculated a set of synthetic spectra
 with only H present (the abundance of all of the other elements were
 set at $-$5.00, while H was at 12.00) while using the relations
 between temperature, density, and radius from the stratified model
 atmospheres. Helium lines were not included because the
 stratification abundances of helium are extremely low. For
 comparison, we also fitted the observed spectra again (only H lines)
 with homogeneous model spectra computed with [M/H] = $+$0.5 and
 ${\log{\frac{n_{\rm He}}{n_{\rm H}}}}$ = $-$3. The results are shown
 in Fig.\,\ref{fig:Teff_g_strat}. The prediction of higher
 spectroscopic gravities by the models including abundance
 stratification shown here was also found in \citet{hulh00} and
 \citet{lehk10}.  The amplitude of the effect might be overestimated
 by the stratified models because in most cases, they predict stronger
 abundance anomalies for blue HB stars than observed (see
 Sect.\,\ref{sec:abu}).

In Fig.\,\ref{fig:Teff_g_strat} all stars move to higher temperatures
and surface gravities when fitted with model spectra from stratified
model atmospheres (as expected from the results of the reverse
  fitting given in Table\,\ref{tab:strat_homo}), but only stars
between 14\,000\,K and 16\,000\,K move closer to the
ZAHB. Cooler stars show too high surface gravities, while hotter ones
move parallel to the TAHB. The principal defect of the
  stratified models is that they generally predict larger abundances
  than observed. In more realistic models, one would expect that the
  shifts shown in Fig.\,\ref{fig:Teff_g_strat} be somewhat
  smaller. The lower temperature stars could then possibly fall
  between the TAHB and ZAHB in the $\log g - T_{\rm eff}$ plane,
which would   give satisfactory results.

\begin{figure}[h!]
\includegraphics[height=1.0\columnwidth,angle=270]{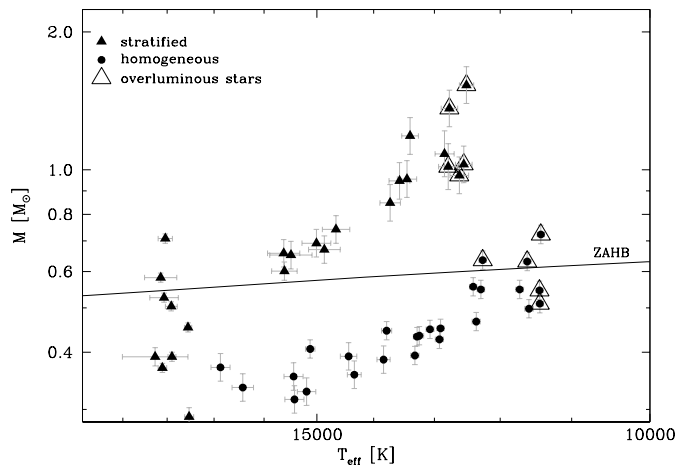}
\caption[]{Masses determined from line profile fits for stars
  bluer than the Grundahl jump, using homogeneous (filled circles) and
  stratified (filled triangles) model spectra.}\label{Fig:mass_strat}
\end{figure}

As a consistency check we derived masses for the new parameters,
adjusting the theoretical $y$-magnitudes of \citet{kuru93} for
[M/H] = $+$0.5 by $-$0\fm3 to account for the stratification
effects. This offset was determined by comparing the fluxes of
stratified and homogeneous model spectra. The results are shown in
Fig.\,\ref{Fig:mass_strat}. Here the values obtained from stratified
model spectra for stars above 14\,000\,K are now closer to the
canonical values than those obtained from homogeneous model
spectra. This provides some support to the notion that the too low
surface gravities and masses found from analyses with homogeneous
model spectra are caused by the mismatch in atmospheric structure between
homogeneous and stratified model atmospheres.
One should keep in mind, however, that the stratified model
atmospheres in general predict much stronger lines than observed,
therefore it is currently unclear whether a fully self-consistent solution can
be achieved. Similarly to the discussion related to the large
  shifts found in Fig.\,\ref{fig:Teff_g_strat}, the masses predicted
  in Fig.\,\ref{Fig:mass_strat} by more realistic stratified models
  for stars just above the temperature threshold of the $u$-jump, in
  which the abundances are less extreme, would be closer to the ZAHB.

\section{Overluminous stars}\label{sec:lum}
\begin{figure}[h!]
\includegraphics[height=1.0\columnwidth,angle=270]{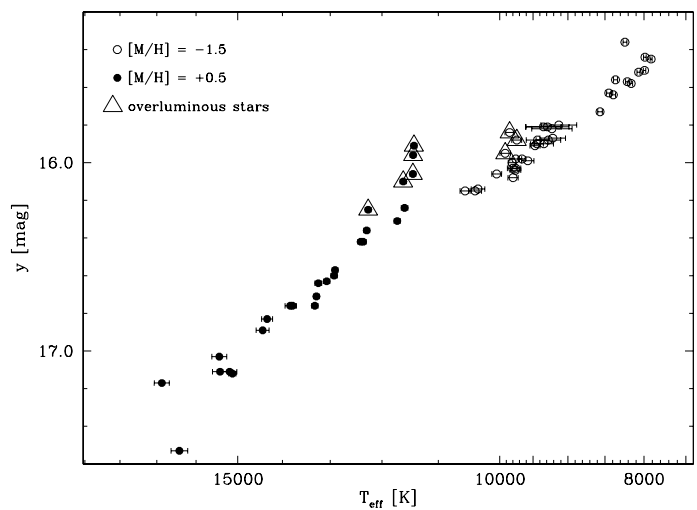}
\caption[]{The $y$ -magnitudes of the HB stars compared with the
  effective temperatures derived from line profile fits.}\label{Fig:Ty}
\end{figure}

One possible explanation for the overluminous stars mentioned in
  Sect.\,\ref{sec:intro} is that they
  have evolved from hotter locations blueward of the $u$-jump and are
  now near the end of their HB phase, evolving towards the asymptotic
  giant branch.  Because the evolution away from the zero-age HB increases
  the {\em overall} luminosity of an HB star, one would expect such
  evolved stars to be overluminous in other bandpasses in addition
to $u$
  compared with stars near the zero-age HB. Figure\,\ref{Fig:Ty} shows the
  relation between the effective temperature derived from line profile
  fits and the $y$ -magnitude of the HB stars. Here especially the hot
  overluminous stars are clearly brighter than the 'normal' stars at
  similar temperatures. To a lesser extent, this is also true for the
  cool overluminous stars. In Figs.\,\ref{fig:Teff_g_hot} and
  \ref{fig:Teff_difflogg}, however, only three of the overluminous
  stars, namely 100 (hot), 103, and 127 (both cool) are clearly
  separated from the majority of the stars by a significantly lower
  surface gravity. It is unclear why the other overluminous stars
  have higher luminosities, because they are inconspicuous with respect to
  the majority of the stars in all other parameters (effective
  temperature, surface gravity, and abundances). However, even three
  stars evolving off the ZAHB in the temperature range 9\,700\,K to
  11\,500\,K pose a problem for evolutionary time-scales, as one would
  expect to find about 100 HB stars close to the ZAHB between
  about 11\,000\,K and 14\,000\,K (corresponding to a range of roughly 0.5
  -- 1.05 in $u-y$ in Fig.\,\ref{fig:cmd_hb}) for each of
  the evolved stars, which is clearly not the case.

If the overluminous stars are post-HB stars evolved from a hotter
location along the blue tail (as suggested by their higher
luminosity), one would expect them to preserve their abundance
anomalies during their evolution towards cooler temperatures
until dredge-up from the deepening hydrogen convection zone, which
causes the surface abundances to return to approximately the normal
cluster abundances. However, such stars are expected to have had a low
He abundance on the HB and a low rotation rate, which may cause them
to maintain their abundance anomalies to temperatures somewhat lower
than the $u$-jump. First, due to the depletion of their surface
helium, the overluminous stars just redward of the $u$-jump may have
slightly shallower convection zones than other stars of the same
effective temperature. However, hydrogen ionization is the main driver
of the convection when the effective temperature is lower than
10\,000\,K and it will lead to progressive helium dredge-up so that
the difference in convection zone depth could be small below
10\,000\,K.  Second, because of the rotation-rate drop in HB stars
blueward of the u-jump, the overluminous stars would be expected to
have low rotation rates if they are evolved from the blue HB.  This,
in turn, would reduce the meridional flows that inhibit diffusion in
stars cooler than the $u$-jump.  However, as may be seen from Fig.\,4
of \citet{qcmr09}, the limiting rotation rate decreases
very rapidly as the effective temperature drops below 10\,000\,K; a
star with a $v_e$ = 8\,km\,s$^{-1}$ would show strong effects of
diffusion above 12\,000\,K but would be mixed below 10\,000\,K.  The
results given in Fig.\,\ref{Fig:abu} indeed show that the effects of
radiative diffusion disappear between the overluminous stars hotter
than the $u$-jump at 11\,500\,K and the overluminous stars cooler than
the $u$-jump at 10\,000\,K.  Accordingly, the low He abundance due to
diffusion and low rotation apparently does not prevent the effects of
diffusion from disappearing rather quickly as a star evolves through
the $u$-jump region to cooler temperatures. This result supports
the general assumption that the abundances of HB stars redward of the
$u$-jump are not affected by diffusion, irrespective of their
evolutionary status. The observations are ahead of the stellar
evolution models in this case. To set the theory on a stronger base,
the evolution calculations with diffusion need to be continued to the
end of the HB; but meridional circulation needs to be included as well
to verify at which effective temperature the mixing occurs and how
this temperature depends on rotation.

\section{Summary and conclusions}
\begin{itemize}
\item
The atmospheric parameters and masses of the hot HB stars in NGC\,288
show a behaviour also seen in other clusters for temperatures between
9\,000\,K and 14\,000\,K. Outside this temperature range, however,
they follow the results found for such stars in $\omega$\,Cen.
\item
The abundances derived from our observations are for most elements
within the abundance range expected from evolutionary models that
include the effects of atomic diffusion and assume a surface mixed
mass of 10$^{-7}$M$_\sun$, as determined previously for sdB stars and
other clusters. The exceptions are helium, which is more deficient
than expected, and phosphorus, which is substantially more abundant
than predicted. The abundances predicted by stratified model
atmospheres, which were not adjusted to observations, are generally
significantly more extreme than observed, except for magnesium.
\item
When the observed spectra were analysed with stratified model spectra,
the HB stars were moved to higher temperatures, surface gravities, and
masses. Since the equilibrium abundances led to excessive adjustments,
more realistic abundance gradients may well lead to models that locate
the HB stars between the TAHB and ZAHB (see
Sect.\,\ref{sec:lineprofstrat}). Model atmospheres including such
improvements are needed to answer this question.
\item
Five of the eight overluminous stars around the $u$-jump that we
observed do not deviate substantially from the other HB stars in the
same temperature range in any of the parameters we determined. We are
therefore at a loss to explain their brighter luminosities. The
remaining three overluminous stars do show lower surface gravities, as
would be expected if they evolved away from the HB. However, they pose
a substantial problem for evolutionary time-scales, because one would
expect to find approximately100 HB stars close to the ZAHB between
about 11\,000\,K and 14\,000\,K (corresponding to a range of roughly
0.5 -- 1.05 in $u-y$ in Fig.\,\ref{fig:cmd_hb}) for each of the
evolved stars, which is clearly not the case.
\item All overluminous stars show the same abundance behaviour as the
majority of the stars in the respective temperature range. This result
supports the general assumption that the abundances of HB stars
redward of the $u$-jump are not affected by diffusion, irrespective of
their evolutionary status.
\end{itemize}

Evolution models including diffusion and stratified model
atmospheres both predict higher-than-observed abundances for many
elements affected by radiative levitation. Evolution models can be
adjusted to reproduce the observed abundances by introducing an ad hoc
defined mixed zone, which is potentially inconsistent with
the observed vertical stratification of at least some elements in the
atmosphere, however. For stratified model atmospheres it is currently
unclear whether a more limited abundance stratification, which would
provide a better description of the observed abundances, would still
be able to explain the photometric anomalies around the $u$-jump. Our results
show definite promise towards solving the long-standing problem of
surface gravity and mass discrepancies for hot HB stars, but  much work is still needed to arrive at a self-consistent solution.

\begin{acknowledgements}
We thank Andr\'e Drews for his careful reduction of the data.  We are grateful to Christian Moni Bidin for sending
his results to allow direct comparisons and for providing a very
thorough and helpful referee report.
\end{acknowledgements}
\bibliography{NGC288}

\begin{thebibliography}{52}
\expandafter\ifx\csname natexlab\endcsname\relax\def\natexlab#1{#1}\fi

\bibitem[{{Behr}(2003)}]{behr03}
{Behr}, B.~B. 2003, \apjs, 149, 67

\bibitem[{{Behr} {et~al.}(2000{\natexlab{a}}){Behr}, {Cohen}, \&
  {McCarthy}}]{beco00}
{Behr}, B.~B., {Cohen}, J.~G., \& {McCarthy}, J.~K. 2000{\natexlab{a}}, \apjl,
  531, L37

\bibitem[{{Behr} {et~al.}(1999){Behr}, {Cohen}, {McCarthy}, \&
  {Djorgovski}}]{beco99}
{Behr}, B.~B., {Cohen}, J.~G., {McCarthy}, J.~K., \& {Djorgovski}, S.~G. 1999,
  \apjl, 517, L135

\bibitem[{{Behr} {et~al.}(2000{\natexlab{b}}){Behr}, {Djorgovski}, {Cohen},
  {McCarthy}, {C{\^o}t{\'e}}, {Piotto}, \& {Zoccali}}]{bedj00}
{Behr}, B.~B., {Djorgovski}, S.~G., {Cohen}, J.~G., {et~al.}
  2000{\natexlab{b}}, \apj, 528, 849

\bibitem[{{Carretta} {et~al.}(2009{\natexlab{a}}){Carretta}, {Bragaglia},
  {Gratton}, {D'Orazi}, \& {Lucatello}}]{cabr09b}
{Carretta}, E., {Bragaglia}, A., {Gratton}, R., {D'Orazi}, V., \& {Lucatello},
  S. 2009{\natexlab{a}}, \aap, 508, 695

\bibitem[{{Carretta} {et~al.}(2011){Carretta}, {Bragaglia}, {Gratton},
  {D'Orazi}, \& {Lucatello}}]{cabr11}
{Carretta}, E., {Bragaglia}, A., {Gratton}, R., {D'Orazi}, V., \& {Lucatello},
  S. 2011, \aap, 535, A121

\bibitem[{{Carretta} {et~al.}(2009{\natexlab{b}}){Carretta}, {Bragaglia},
  {Gratton}, \& {Lucatello}}]{cabr09a}
{Carretta}, E., {Bragaglia}, A., {Gratton}, R., \& {Lucatello}, S.
  2009{\natexlab{b}}, \aap, 505, 139

\bibitem[{{Carretta} {et~al.}(2000){Carretta}, {Gratton}, {Clementini}, \&
  {Fusi Pecci}}]{cagr00}
{Carretta}, E., {Gratton}, R.~G., {Clementini}, G., \& {Fusi Pecci}, F. 2000,
  \apj, 533, 215

\bibitem[{{Crocker} {et~al.}(1988){Crocker}, {Rood}, \& {O'Connell}}]{crro88}
{Crocker}, D.~A., {Rood}, R.~T., \& {O'Connell}, R.~W. 1988, \apj, 332, 236

\bibitem[{{Drews}(2005)}]{drews2005}
{Drews}, A. 2005, Diploma Thesis Christian-Albrechts Universit\"at Kiel

\bibitem[{{Fabbian} {et~al.}(2005){Fabbian}, {Recio-Blanco}, {Gratton}, \&
  {Piotto}}]{fare05}
{Fabbian}, D., {Recio-Blanco}, A., {Gratton}, R.~G., \& {Piotto}, G. 2005,
  \aap, 434, 235

\bibitem[{{Ferraro} {et~al.}(1998){Ferraro}, {Paltrinieri}, {Pecci}, {Rood}, \&
  {Dorman}}]{fpfd98}
{Ferraro}, F.~R., {Paltrinieri}, B., {Pecci}, F.~F., {Rood}, R.~T., \&
  {Dorman}, B. 1998, \apj, 500, 311

\bibitem[{{Gratton} {et~al.}(2010){Gratton}, {Carretta}, {Bragaglia},
  {Lucatello}, \& {D'Orazi}}]{grca10}
{Gratton}, R.~G., {Carretta}, E., {Bragaglia}, A., {Lucatello}, S., \&
  {D'Orazi}, V. 2010, \aap, 517, A81

\bibitem[{{Grundahl} {et~al.}(1999){Grundahl}, {Catelan}, {Landsman},
  {Stetson}, \& {Andersen}}]{grca99}
{Grundahl}, F., {Catelan}, M., {Landsman}, W.~B., {Stetson}, P.~B., \&
  {Andersen}, M.~I. 1999, \apj, 524, 242

\bibitem[{{Harris}(1996)}]{harr96}
{Harris}, W.~E. 1996, \aj, 112, 1487 (version Dec. 2010)

\bibitem[{{Horne}(1986)}]{horn86}
{Horne}, K. 1986, \pasp, 98, 609

\bibitem[{{Hui-Bon-Hoa} {et~al.}(2000){Hui-Bon-Hoa}, {LeBlanc}, \&
  {Hauschildt}}]{hulh00}
{Hui-Bon-Hoa}, A., {LeBlanc}, F., \& {Hauschildt}, P.~H. 2000, \apjl, 535, L43

\bibitem[{{Khalack} {et~al.}(2010){Khalack}, {LeBlanc}, \& {Behr}}]{khlb10}
{Khalack}, V., {LeBlanc}, F., \& {Behr}, B.~B. 2010, \mnras, 407, 1767

\bibitem[{{Khalack} {et~al.}(2008){Khalack}, {LeBlanc}, {Behr}, {Wade}, \&
  {Bohlender}}]{klbw08}
{Khalack}, V.~R., {LeBlanc}, F., {Behr}, B.~B., {Wade}, G.~A., \& {Bohlender},
  D. 2008, \aap, 477, 641

\bibitem[{{Khalack} {et~al.}(2007){Khalack}, {LeBlanc}, {Bohlender}, {Wade}, \&
  {Behr}}]{klbw07}
{Khalack}, V.~R., {LeBlanc}, F., {Bohlender}, D., {Wade}, G.~A., \& {Behr},
  B.~B. 2007, \aap, 466, 667

\bibitem[{{Kurucz}(1993)}]{kuru93}
{Kurucz}, R.~L. 1993, ATLAS9 Stellar Atmospheres Program and 2 km s$^{-1}$
  grid, CR-ROM No. 13

\bibitem[{{Lane} {et~al.}(2010){Lane}, {Kiss}, {Lewis}, {Ibata}, {Siebert},
  {Bedding}, {Sz{\'e}kely}, {Balog}, \& {Szab{\'o}}}]{laki10}
{Lane}, R.~R., {Kiss}, L.~L., {Lewis}, G.~F., {et~al.} 2010, \mnras, 406, 2732

\bibitem[{{LeBlanc} {et~al.}(2010){LeBlanc}, {Hui-Bon-Hoa}, \&
  {Khalack}}]{lehk10}
{LeBlanc}, F., {Hui-Bon-Hoa}, A., \& {Khalack}, V.~R. 2010, \mnras, 409, 1606

\bibitem[{{LeBlanc} {et~al.}(2009){LeBlanc}, {Monin}, {Hui-Bon-Hoa}, \&
  {Hauschildt}}]{lmhh09}
{LeBlanc}, F., {Monin}, D., {Hui-Bon-Hoa}, A., \& {Hauschildt}, P.~H. 2009,
  \aap, 495, 937

\bibitem[{{Michaud} {et~al.}(2008){Michaud}, {Richer}, \& {Richard}}]{miri08}
{Michaud}, G., {Richer}, J., \& {Richard}, O. 2008, \apj, 675, 1223

\bibitem[{{Michaud} {et~al.}(2011){Michaud}, {Richer}, \& {Richard}}]{miri11}
{Michaud}, G., {Richer}, J., \& {Richard}, O. 2011, \aap, 529, A60

\bibitem[{{Michaud} {et~al.}(1983){Michaud}, {Vauclair}, \&
  {Vauclair}}]{miva83}
{Michaud}, G., {Vauclair}, G., \& {Vauclair}, S. 1983, \apj, 267, 256

\bibitem[{{Moehler}(2001)}]{moeh01}
{Moehler}, S. 2001, \pasp, 113, 1162

\bibitem[{{Moehler} {et~al.}(2011){Moehler}, {Dreizler}, {Lanz}, {Bono},
  {Sweigart}, {Calamida}, \& {Nonino}}]{modr11}
{Moehler}, S., {Dreizler}, S., {Lanz}, T., {et~al.} 2011, \aap, 526, A136

\bibitem[{{Moehler} {et~al.}(1995){Moehler}, {Heber}, \& {de Boer}}]{mohe95}
{Moehler}, S., {Heber}, U., \& {de Boer}, K.~S. 1995, \aap, 294, 65

\bibitem[{{Moehler} {et~al.}(1997){Moehler}, {Heber}, \& {Rupprecht}}]{mohe97}
{Moehler}, S., {Heber}, U., \& {Rupprecht}, G. 1997, \aap, 319, 109

\bibitem[{{Moehler} {et~al.}(2003){Moehler}, {Landsman}, {Sweigart}, \&
  {Grundahl}}]{mola03}
{Moehler}, S., {Landsman}, W.~B., {Sweigart}, A.~V., \& {Grundahl}, F. 2003,
  \aap, 405, 135

\bibitem[{{Moehler} {et~al.}(2000){Moehler}, {Sweigart}, {Landsman}, \&
  {Heber}}]{mosw00}
{Moehler}, S., {Sweigart}, A.~V., {Landsman}, W.~B., \& {Heber}, U. 2000, \aap,
  360, 120

\bibitem[{{Monelli} {et~al.}(2013){Monelli}, {Milone}, {Stetson}, {Marino},
  {Cassisi}, {del Pino Molina}, {Salaris}, {Aparicio}, {Asplund}, {Grundahl},
  {Piotto}, {Weiss}, {Carrera}, {Cebri{\'a}n}, {Murabito}, {Pietrinferni}, \&
  {Sbordone}}]{momi13}
{Monelli}, M., {Milone}, A.~P., {Stetson}, P.~B., {et~al.} 2013, \mnras, 431,
  2126

\bibitem[{{Moni Bidin} {et~al.}(2007){Moni Bidin}, {Moehler}, {Piotto},
  {Momany}, \& {Recio-Blanco}}]{momo07}
{Moni Bidin}, C., {Moehler}, S., {Piotto}, G., {Momany}, Y., \& {Recio-Blanco},
  A. 2007, \aap, 474, 505

\bibitem[{{Moni Bidin} {et~al.}(2009){Moni Bidin}, {Moehler}, {Piotto},
  {Momany}, \& {Recio-Blanco}}]{momo09}
{Moni Bidin}, C., {Moehler}, S., {Piotto}, G., {Momany}, Y., \& {Recio-Blanco},
  A. 2009, \aap, 498, 737

\bibitem[{{Moni Bidin} {et~al.}(2012){Moni Bidin}, {Villanova}, {Piotto},
  {Moehler}, {Cassisi}, \& {Momany}}]{movi12}
{Moni Bidin}, C., {Villanova}, S., {Piotto}, G., {et~al.} 2012, \aap, 547, A109

\bibitem[{{Moni Bidin} {et~al.}(2011){Moni Bidin}, {Villanova}, {Piotto},
  {Moehler}, \& {D'Antona}}]{movi11}
{Moni Bidin}, C., {Villanova}, S., {Piotto}, G., {Moehler}, S., \& {D'Antona},
  F. 2011, \apjl, 738, L10

\bibitem[{{Moon} \& {Dworetsky}(1985)}]{modw85}
{Moon}, T.~T. \& {Dworetsky}, M.~M. 1985, \mnras, 217, 305

\bibitem[{{Napiwotzki}(1997)}]{napi97}
{Napiwotzki}, R. 1997, \aap, 322, 256

\bibitem[{{Napiwotzki} {et~al.}(2004){Napiwotzki}, {Yungelson}, {Nelemans},
  {Marsh}, {Leibundgut}, {Renzini}, {Homeier}, {Koester}, {Moehler},
  {Christlieb}, {Reimers}, {Drechsel}, {Heber}, {Karl}, \& {Pauli}}]{nayu04}
{Napiwotzki}, R., {Yungelson}, L., {Nelemans}, G., {et~al.} 2004, in
  Astronomical Society of the Pacific Conference Series, Vol. 318,
  Spectroscopically and Spatially Resolving the Components of the Close Binary
  Stars, ed. R.~W. {Hilditch}, H.~{Hensberge}, \& K.~{Pavlovski}, 402--410

\bibitem[{{Pace} {et~al.}(2006){Pace}, {Recio-Blanco}, {Piotto}, \&
  {Momany}}]{pare06}
{Pace}, G., {Recio-Blanco}, A., {Piotto}, G., \& {Momany}, Y. 2006, \aap, 452,
  493

\bibitem[{{Pasquini} {et~al.}(2000){Pasquini}, {Avila}, {Allaert}, {Ballester},
  {Biereichel}, {Buzzoni}, {Cavadore}, {Dekker}, {Delabre}, {Ferraro}, {Hill},
  {Kaufer}, {Kotzlowski}, {Lizon}, {Longinotti}, {Moureau}, {Palsa}, \&
  {Zaggia}}]{pasq00}
{Pasquini}, L., {Avila}, G., {Allaert}, E., {et~al.} 2000, in Society of
  Photo-Optical Instrumentation Engineers (SPIE) Conference Series, Vol. 4008,
  Optical and IR Telescope Instrumentation and Detectors, ed. M.~{Iye} \& A.~F.
  {Moorwood}, 129--140

\bibitem[{{Peterson} {et~al.}(1995){Peterson}, {Rood}, \& {Crocker}}]{perc95}
{Peterson}, R.~C., {Rood}, R.~T., \& {Crocker}, D.~A. 1995, \apj, 453, 214

\bibitem[{{Piotto} {et~al.}(2013){Piotto}, {Milone}, {Marino}, {Bedin},
  {Anderson}, {Jerjen}, {Bellini}, \& {Cassisi}}]{pimi13}
{Piotto}, G., {Milone}, A.~P., {Marino}, A.~F., {et~al.} 2013, \apj, 775, 15

\bibitem[{{Quievy} {et~al.}(2009){Quievy}, {Charbonneau}, {Michaud}, \&
  {Richer}}]{qcmr09}
{Quievy}, D., {Charbonneau}, P., {Michaud}, G., \& {Richer}, J. 2009, \aap,
  500, 1163

\bibitem[{{Roh} {et~al.}(2011){Roh}, {Lee}, {Joo}, {Han}, {Sohn}, \&
  {Lee}}]{role11}
{Roh}, D.-G., {Lee}, Y.-W., {Joo}, S.-J., {et~al.} 2011, \apjl, 733, L45

\bibitem[{{Salgado} {et~al.}(2013){Salgado}, {Moni Bidin}, {Villanova},
  {Geisler}, \& {Catelan}}]{samo13}
{Salgado}, C., {Moni Bidin}, C., {Villanova}, S., {Geisler}, D., \& {Catelan},
  M. 2013, \aap, 559, A101

\bibitem[{{Sz{\'e}kely} {et~al.}(2007){Sz{\'e}kely}, {Kiss}, {Szatm{\'a}ry},
  {Cs{\'a}k}, {Bakos}, \& {Bedding}}]{szki07}
{Sz{\'e}kely}, P., {Kiss}, L.~L., {Szatm{\'a}ry}, K., {et~al.} 2007,
  Astronomische Nachrichten, 328, 879

\bibitem[{{ten Bruggencate}(1927)}]{ten27}
{ten Bruggencate}, P. 1927, Sternhaufen (Berlin: Springer)

\bibitem[{{Tonry} \& {Davis}(1979)}]{toda79}
{Tonry}, J. \& {Davis}, M. 1979, \aj, 84, 1511

\bibitem[{{Yong} {et~al.}(2008){Yong}, {Grundahl}, {Johnson}, \&
  {Asplund}}]{yogr08}
{Yong}, D., {Grundahl}, F., {Johnson}, J.~A., \& {Asplund}, M. 2008, \apj, 684,
  1159

\end{thebibliography}
\end{document}